\documentclass[%
reprint,
floatfix,
superscriptaddress,
amsmath,amssymb,
aps,
pra,
]{revtex4-2}
\usepackage[utf8]{inputenc}
\usepackage[english]{babel}
\usepackage{blindtext}
\usepackage{amsmath}
\usepackage{xcolor,colortbl}
\definecolor{woNbPad}{RGB}{240,142,80}
\definecolor{wNbPad}{RGB}{199,0,199}
\usepackage{array}
\newcommand\Tstrut{\rule{0pt}{2.15ex}}         
\newcommand\Bstrut{\rule[-0.45ex]{0pt}{0pt}}   
\newcommand\Tstruttop{\rule{0pt}{2.5ex}}         
\newcommand\Bstruttop{\rule[-0.6ex]{0pt}{0pt}}   
\newcolumntype{C}{>{$\displaystyle}c<{$}}
\usepackage{graphicx}   
\graphicspath{{figs/}}   
\usepackage{dcolumn}
\usepackage{bm}
\usepackage[hidelinks]{hyperref}
\usepackage{epstopdf}
\usepackage{braket}
\usepackage{upgreek}
\usepackage{amsmath}
\usepackage{siunitx}
\usepackage{changes}
\usepackage{natbib}
\usepackage[title]{appendix}
\usepackage[all]{hypcap}
\usepackage[capitalise,nameinlink]{cleveref}

\usepackage{multirow}

\newcommand*{\scrpt}[1]{\mathrm{#1}}
\newcommand*{\s}[1]{\ensuremath{_\scrpt{#1}}}

\usepackage{cleveref} 
\crefname{section}{Appendix}{Appendices} 

\begin{document}
\title{Double-Pumped~Kerr~Parametric~Amplifier~Beyond~the~Gain-Bandwidth~Limit}

\author{Nicolas~Zapata}
\email{Both authors contributed equally}
\affiliation{Institute for Quantum Materials and Technology, Karlsruhe Institute of Technology, 76131 Karlsruhe, Germany}

\author{Najmeh~Etehadi~Abari}
\email{Both authors contributed equally}
\affiliation{Institute for Quantum Materials and Technology, Karlsruhe Institute of Technology, 76131 Karlsruhe, Germany}

\author{Mitchell~Field}
\affiliation{Institute for Quantum Materials and Technology, Karlsruhe Institute of Technology, 76131 Karlsruhe, Germany}

\author{Patrick~Winkel}
\email{current affiliation: Alice \& Bob, 49 boulevard du G\'en\'eral Martial Valin, 75015 Paris, France}
\affiliation{Institute for Quantum Materials and Technology, Karlsruhe Institute of Technology, 76131 Karlsruhe, Germany}

\author{Simon~Geisert}
\affiliation{Institute for Quantum Materials and Technology, Karlsruhe Institute of Technology, 76131 Karlsruhe, Germany}

\author{Soeren~Ihssen}
\affiliation{Institute for Quantum Materials and Technology, Karlsruhe Institute of Technology, 76131 Karlsruhe, Germany}

\author{Anja~Metelmann}
\affiliation{Institute for Quantum Materials and Technology, Karlsruhe Institute of Technology, 76131 Karlsruhe, Germany}
\affiliation{Institute for Theory of Condensed Matter, Karlsruhe Institute of Technology, 76131 Karlsruhe, Germany}
\affiliation{Institut de Science et d’Ingénierie Supramoléculaires (ISIS, UMR7006), University of Strasbourg and CNRS, 67000 Strasbourg, France
}

\author{Ioan~M.~Pop}
\email{ioan.pop@kit.edu}
\affiliation{Institute for Quantum Materials and Technology, Karlsruhe Institute of Technology, 76131 Karlsruhe, Germany}
\affiliation{Physikalisches Institut, Karlsruhe Institute of Technology, 76131 Karlsruhe, Germany}
\affiliation{Physics~Institute~1,~Stuttgart~University,~70569~Stuttgart,~Germany}

\date{\today}

\begin{abstract}
Superconducting standing-wave parametric amplifiers are crucial for the readout of microwave quantum devices. Despite significant improvements in recent years, the need to operate near an instability point imposes a fundamental constraint: the instantaneous bandwidth decreases with increasing amplifier gain. Here we show that it is possible to obtain parametric amplification without instability by using two simultaneous drives that activate phase-preserving gain and frequency conversion. Realized in a granular aluminum dimer with Kerr nonlinearity, our method demonstrates~a sixfold bandwidth increase at 20~dB gain, surpasses the conventional gain–bandwidth scaling up~to~25~dB, and remains near the quantum limit.
\end{abstract}



\maketitle

The remarkable noise properties of superconducting parametric amplifiers have promoted their widespread integration in systems requiring high fidelity readout of quantum devices \cite{Vijay2011Mar,Roy2016Aug,Aumentado2020Jul}. They are classified as standing-wave~\cite{Roy2016Aug,Aumentado2020Jul} or traveling-wave parametric amplifiers \cite{HoEom2012Aug,Macklin2015Sep,Malnou2021Jan,Esposito2021Sep}, depending on whether amplification occurs within a superconducting cavity or along a nonlinear transmission line, respectively. The forte of standing-wave parametric amplifiers lies in their ability to deliver high gain and quantum-limited noise performance with flexible engineering and fabrication~\cite{Aumentado2020Jul,Sivak2020,Winkel2020DJA,Xu2023Feb,Kaufman2025Jul}. However, they need to operate near an instability point, which results in a decrease of the instantaneous bandwidth (BW) with increasing amplifier gain \cite{Roy2016Aug,Clerk2010Apr}, commonly referred to as the gain–bandwidth tradeoff (GBW). 

In practice, the operational BW of standing-wave amplifiers is restricted by the GBW tradeoff to the tens of~MHz range at $\sim$~20~dB gain levels, needed for low-noise measurements. This limits their applicability in systems requiring bandwidths close to the hundreds of MHz range, such as frequency-multiplexed qubit readout \cite{Heinsoo2018Sep,Krinner2022May,White2023Jan,BibEntry2023Feb}, broadband quantum optics experiments in the microwave domain \cite{Esposito2022Apr,Agusti2022Jun,Qiu2023May}, and dark-matter searches \cite{Lamoreaux2013Aug,Backes2021Feb,Braggio2022Sep}.
While state-of-the-art impedance engineering techniques provide a tenfold improvement in BW for gain levels exceeding 20~dB \cite{Mutus2014Jun,Roy2015Dec,Naaman2022May,Grebel2021Apr,Ezenkova2022Dec,Kaufman2023Nov,Zhou2025Mar,Hung2025Apr,Joshi2025Oct}, fulfilling a long-standing need in the community, they come at the cost of stringent fabrication requirements and fixed gain point operation. Deviations in the impedance-matching network can degrade the device BW and adjusting the gain to different values can introduce unwanted variations in the gain profile reminiscent of the GBW tradeoff \cite{Mutus2014Jun,Ezenkova2022Dec,Zhou2025Mar}.

Ideally, the amplifier should operate away from the instability point in order to avoid the GBW limit. This can be achieved using a complementary strategy which consists in applying multiple pumps that simultaneously enable frequency conversion and parametric amplification \cite{Metelmann2014Apr,Nunnenkamp2014Jul,Metelmann2015Jun,Metelmann2022Bog,Ruddy2024Aug}. When both processes are optimally balanced, they provide dynamical stability to the system and enable the amplifier to overcome the inherent GBW limit. A key advantage is that the bandwidth can be optimized in-situ for any gain level by adjusting the frequency and power of the pump tones. 

Amplifiers operating away from instability have been successfully realized in both superconducting circuits~\cite{Chien2020Apr,Jiang2023Apr,Dassonneville2021May} and electromechanical platforms~\cite{Ockeloen-Korppi2016Oct,Toth2017Aug}. Electromechanical systems have demonstrated gain levels up to 40~dB without a GBW limitation. However, the device BW remained limited to the~kHz range, well below that required for quantum device readout. In contrast, superconducting circuit implementations based on three-wave mixing of a Josephson Parametric Converter have reported an improved GBW product up to 15 dB gain, with BWs in the tens of MHz range~\cite{Chien2020Apr}. Yet, performance at higher gain was limited by residual Kerr effects and higher-order nonlinearities \cite{Chien2020Apr,Jiang2023Apr,Dassonneville2021May}.

\begin{figure}[!t]
\includegraphics[width = 1\columnwidth]{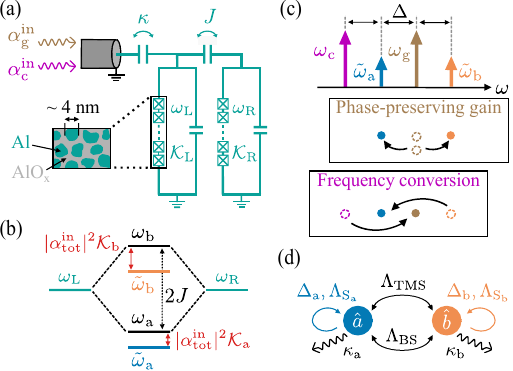}
\caption{\textbf{Device design and modes of operation.} \textbf{(a)}~Circuit diagram of the granular aluminum parametric amplifier (grAlPA). The Bose-Hubbard dimer design~\cite{Eichler2014Dimer, Winkel2020DJA} is similar to Ref.~\cite{Zapata2024Dec} and consists of a pair of capacitively coupled grAl resonators with frequencies and self-Kerr coefficients denoted by $\omega\mathrm{_i}$ and $\mathcal{K}\mathrm{_{i}}$ (i~=~L,~R), respectively. The nonlinearity of grAl resonators can be modeled as an effective Josephson junction array \cite{Maleeva2018}. Resonator L is coupled with rate $\kappa$ to an input port, through which two pump tones are applied to activate gain (brown) and frequency conversion (purple). \textbf{(b)}~Dressed mode structure of the dimer. The hopping interaction $J$ gives rise to hybridized modes $\omega \mathrm{_{a/b}}$. When driven, both modes are red shifted by $|\alpha\mathrm{^{in}_{tot}}|^{2} \mathcal{K}\mathrm{_{a/b}}$ to $\tilde{\omega}\mathrm{_{a/b}}$, where $|\alpha\mathrm{^{in}_{tot}}|$ is proportional to the total input power and $\mathcal{K}\mathrm{_{a/b}}$ are effective Kerr coefficients (see \cref{A_Theory}). 
\textbf{(c)}~Configuration for double-pumping experiments. A pump applied at $\omega\mathrm{_g}$~=~$(\bar{\omega}\mathrm{_a}+\bar{\omega}\mathrm{_b})/2$, produce phase-preserving gain by transferring two pump photons into one signal and one idler photon split between the dimer modes. A second pump, applied at $\omega\mathrm{_c}$~=~$(\omega\mathrm{_g}+\bar{\omega}\mathrm{_a}-\bar{\omega}\mathrm{_b})$, converts a photons between $\bar{\omega}\mathrm{_a}$ and $\bar{\omega}\mathrm{_b}$, mediated by the creation of a gain pump photon. 
\textbf{(d)}~Dynamics of the system following \cref{eq_linearized_H}. Each mode has a damping rate $\kappa\mathrm{_j}$, a frequency detuning $\Delta\mathrm{_j}$ relative to the rotating frame at $\omega\mathrm{_g}$, and single-mode squeezing interactions $\Lambda\mathrm{_{S_{j}}}$ (j~=~a,~b). The modes are coupled through beam-splitter $\Lambda\mathrm{_{BS}}$ and two-mode squeezing $\Lambda\mathrm{_{TMS}}$ interactions, activated by the parametric drives.}
\label{fig_design}
\end{figure}
In this work, we present a device that relies solely on its Kerr nonlinearity to realize a parametric amplifier with a nonconventional GBW product. The device, referred to as grAlPA \cite{Zapata2024Dec}, is implemented in the form of a lumped-element granular aluminum (grAl) Bose-Hubbard dimer, as shown in \cref{fig_design}(a). We exploit the reduced higher order nonlinearities of grAl and the absence of harmonics $\lesssim$~15~GHz to achieve a sixfold BW improvement for 20~dB gain, with an enhanced scaling and near quantum limited noise performance up to 25~dB. This performance demonstrates the potential of multi-pump parametric devices and recommends grAl as a source of pure four-wave mixing nonlinearity.

The equivalent grAlPA circuit is shown in~\cref{fig_design}(a), which consists of two lumped-element grAl resonators with frequencies $\omega\mathrm{_{L/R}}$ coupled through a capacitive interaction $J$. As depicted in \cref{fig_design}(b), the two resonators hybridize and form a pair of dimer modes with frequencies $\omega\mathrm{_{a/b}}$ (see~\cref{A_Theory} and \cref{A_circuitParameters}). We probe the device in reflection through a single microwave port coupled to the left resonator, resulting in a total coupling strength 
$\kappa$, which is shared between the modes. Single-tone spectroscopy data (see~\cref{A_circuitParameters}) shows negligible internal losses in both resonators, such that the dissipation of each mode is dominated by coupling to the measurement port. Both the inductance and nonlinearity of each resonator originate from a \SI[parse-numbers=false]{7\times0.2\times0.04}{\micro m} grAl strip with resistivity 830~\SI{}{\micro \ohm}cm \cite{Zapata2024Dec}, which we model as an array of Josephson junctions exhibiting self-Kerr nonlinearities $\mathcal{K}\mathrm{_{L/R}}$ \cite{Maleeva2018}. Notably, the granular structure enables the implementation of an effective array of $\sim$~10$\mathrm{^3}$ junctions in the volume of the strip, which dilutes higher-order nonlinearities by three orders of magnitude compared to $\mathcal{K}\mathrm{_{L/R}}$. Moreover, the absence of resonator harmonics in the frequency vicinity of the dimer modes, avoids spurious cross-Kerr interactions during pumping.
\begin{figure}[!t]
\includegraphics[width = 1\columnwidth]{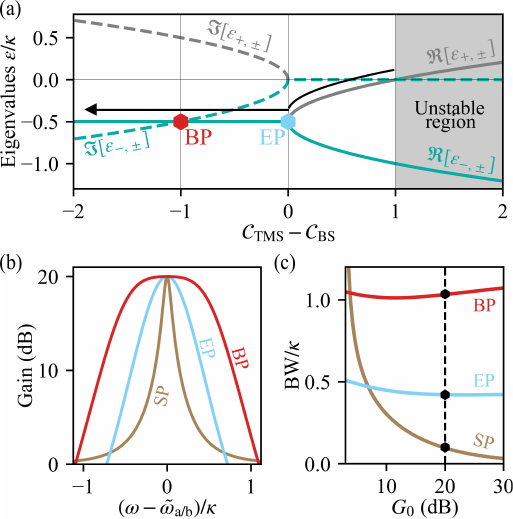}
\caption{\textbf{Eigenvalues and operational modes of a symmetric grAlPA with balanced single-mode squeezing interactions ($|\Delta\mathrm{_{a,b}}|=2|\Lambda\mathrm{_{S}}|$).} \textbf{(a)}~Real and imaginary parts of the eigenvalues $\epsilon\mathrm{_{\pm,\pm}}$ (cf. \cref{Eigenvalues_MXP}) vs. cooperativity difference $\mathcal{C\mathrm{_{TMS}}}-\mathcal{C\mathrm{_{BS}}}$. When the two cooperativities are equal the eigenvalues become degenerate and the system shows an exceptional point (EP). For optimally imbalance $\mathcal{C\mathrm{_{TMS}}}-\mathcal{C\mathrm{_{BS}}}=-1$, the system reaches the Bogoliubov point (BP), where the gain profile exhibits a flattened maximum. The grAlPA surpasses the GBW limit when operated at any point between the EP and BP. Below the BP, the gain profiles split into two peaks. Above the EP, the system provides gain as it approaches the instability region $\mathcal{C\mathrm{_{TMS}}}-\mathcal{C\mathrm{_{BS}}}>1$, exhibiting a conventional GBW scaling. Moving along the black arrow, the amplifier BW gradually increases. \textbf{(b)}~Gain profiles for three possible operational modes of the grAlPA: single-pump (SP), double-pump at exceptional point (EP), and double-pump at the Bogoliubov point (BP). All curves are calculated in the quadrature representation of the hybridized basis (see \cref{A_Theory}). \textbf{(c)}~Bandwidth scaling vs maximum gain $G\mathrm{_0}$ for the modes of operation in panel (b). Both the EP and BP regimes overcome the SP GBW product.}
\label{fig_phase_diagram}
\end{figure}

Driving the device with multiple pumps enables simultaneous four-wave mixing processes, as depicted in~\cref{fig_design}(c).~A~pump~applied~between the Kerr-shifted modes $\tilde{\omega}\mathrm{_{a,b}}$ at frequency~$\omega\mathrm{_g}$~=~$(\tilde{\omega}\mathrm{_a}+\tilde{\omega}\mathrm{_b})/2$, which we call the gain pump, generates phase-preserving gain in which two pump photons transform into one signal and one idler photon, split between the dimer modes. The application of a second pump~at frequency $\omega\mathrm{_c}=\omega\mathrm{_g}+\tilde{\omega}\mathrm{_a}-\tilde{\omega}\mathrm{_b}$, called the conversion pump, activates frequency conversion between photons at $\tilde{\omega}\mathrm{_a}$ and $\tilde{\omega}\mathrm{_b}$, mediated by photons from the gain pump. Note that, a device using four-wave-mixing also enables frequency conversion using a single pump at $(\tilde{\omega}\mathrm{_b}-\tilde{\omega}\mathrm{_a})/2$. For our current device, this frequency is in the~100 MHz range, outside of the setup bandwidth (see~\cref{A_Setup}).

The effective dynamics of the system, expressed in the hybridized basis and in a frame rotating at frequency~$\omega\mathrm{_g}$, is illustrated in the mode diagram of~\cref{fig_design}(d)
and described by the Hamiltonian
\begin{equation}
\begin{aligned}
\hat{H}\s{H}/\hbar  & = -\sum_{i= \mathrm{a,b}} \left[\Delta_i \,\hat c^{\dagger}_{i} \hat c_{i} - \left(\Lambda_{\mathrm{S}_{i}} \, \hat c^{\dagger}_{i} \hat c^{\dagger}_{i} + \mathrm{h.c.} \right) \right]\\
 & + \Big( \Lambda\mathrm{_{TMS}} \, \hat c^{\dagger}_{\mathrm{a}} \hat c^{\dagger}_{\mathrm{b}} + \mathrm{h.c.} \Big) +
\Big( \Lambda\mathrm{_{BS}} \,\hat c^{\dagger}_{\mathrm{a}} \hat c_{\mathrm{b}} + \mathrm{h.c.} \Big),
\end{aligned}
\label{eq_linearized_H}
\end{equation}
where $c\mathrm{_{a/b}}$ corresponds to the linearized bosonic dimer modes and $\Delta\mathrm{_{a/b}}$~=~$\omega\mathrm{_g}-\tilde{\omega}\mathrm{_{a/b}}$ denotes their respective frequency detuning. The remaining terms in~\cref{eq_linearized_H}, correspond to single-mode squeezing, two-mode squeezing (TMS), and beam splitter interactions (BS) with coupling strengths $|\Lambda\mathrm{_{S_{a/b}}}|$, $|\Lambda\mathrm{_{TMS}}|$ and $|\Lambda\mathrm{_{BS}}|$, respectively. 

Tuning the system to a time-independent regime allows fast-rotating terms to be neglected, and the exact pumping configuration determines which coupling parameters dominate. During single-pump operation, with the gain pump at frequency $\omega\mathrm{_g}$~=~$(\tilde{\omega}\mathrm{_a}+\tilde{\omega}\mathrm{_b})/2$, the dynamics are governed primarily by two-mode squeezing interactions, effectively implementing a phase-preserving amplifier with a standard GBW limit.
When both gain and conversion pumps are applied, all interaction terms in~\cref{eq_linearized_H} become relevant. 
Two-mode squeezing $\Lambda_{\mathrm{TMS}}$ and single-mode squeezing $\Lambda_{\mathrm{S}_{\mathrm{a/b}}}$ amplify the signal, but also push the system towards instability. Similar to Refs.~\cite{Ruddy2024Aug,Jiang2023Apr}, we counterbalance the effect of single-mode squeezing by adjusting the detuning terms in Hamiltonian~(\ref{eq_linearized_H}), such that $\Delta_a = - \Delta_b=\Delta$, $\Lambda_{\mathrm{S}_a} = \Lambda_{\mathrm{S}_b} = \Lambda_{\mathrm{S}}$ and  $|\Delta_{\mathrm{a}}| = |\Delta_{\mathrm{b}}| = 2 |\Lambda_{\mathrm{S}}|$. We then adjust the beam-splitter interaction to counteract the instability induced by two-mode squeezing, ensuring that the real parts of the dynamical matrix eigenvalues remain negative (see~\cref{A_Theory}) and anchoring the system in the stable regime.

In order to enable a tractable analytical study, we model an idealized Bose-Hubbard dimer with a symmetric configuration where each resonator is coupled to a different microwave port with the same damping rate $\kappa$. 
For the real device of~\cref{fig_design}(a), the reasoning remains valid but the gain curves have to be calculated numerically. When the single-mode squeezing terms are optimally balanced, the eigenvalues take the simplified form
\begin{align}
\label{Eigenvalues_MXP}
\epsilon_{\pm, \pm} & = \frac{\kappa}{2} \left(-1 \pm \sqrt{\mathcal{C}_{\mathrm{TMS}} - \mathcal{C}_{\mathrm{BS}}} \right), 
\end{align}
where $\mathcal{C}_{\alpha} = 4 |\Lambda_{\alpha}|^2/ \kappa^2$ with $\alpha = \mathrm{TMS}, \mathrm{BS}, \mathrm{S} $, represent the cooperativities for the corresponding interactions.

\cref{fig_phase_diagram}(a) illustrates the normalized eigenvalues of the system dynamics, which remain stable as long as $\mathcal{C}_{\mathrm{TMS}} - \mathcal{C}_{\mathrm{BS}} < 1$. 
Within this region, two distinct operating points emerge. 
The exceptional point (EP) is reached when $\mathcal{C}\s{TMS} = \mathcal{C}\s{BS}$, in which case all eigenvalues coalesce into a single real value $\epsilon_{\pm, \pm}|_{\mathrm{EP}} \rightarrow - \kappa /2$. 
The Bogoliubov point (BP) arises when the BS cooperativity surpasses the TMS cooperativity such that the imaginary part of the eigenvalues~(cf.~\cref{Eigenvalues_MXP}) become equal to their real part, corresponding to $\mathcal{C}_{\mathrm{TMS}} - \mathcal{C}_{\mathrm{BS}} = -1$. At this point, the Hamiltonian in~\cref{eq_linearized_H} can be diagonalized via Bogoliubov transformations of the bosonic modes, and the system dynamics map onto those of a pair of Bogoliubov modes coupled by a hopping interaction, as discussed by~Ref.~\cite{Metelmann2022Bog}.

\begin{figure*}[t!]
\includegraphics[width=6.67in]{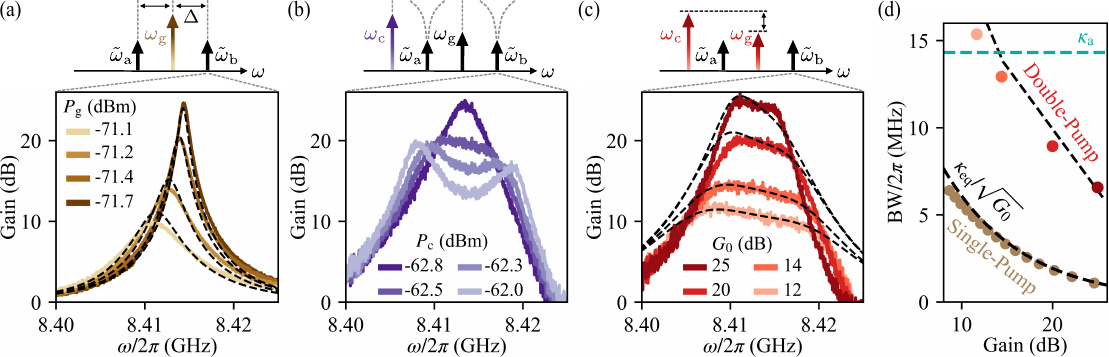}
\caption{\textbf{Improved gain-bandwidth scaling close to the BP.} In panel \textbf{(a)} we show the gain performance under a single pump, in panel \textbf{(b)} we show the two-pump protocol that enables bandwidth optimization, and in panel \textbf{(c)} we demonstrate the optimized operation of the amplifier close to the BP. The schematics above each panel depict the corresponding pump configurations. As visible in panel (a), the gain pump with power $P\mathrm{_g}$ applied at frequency $\omega\mathrm{_g}$~=~$\tilde{\omega}\mathrm{_b}$~-~$\Delta$~=~($\tilde{\omega}\mathrm{_a}$~+~$\tilde{\omega}\mathrm{_b}$)$/2$ gives rise to phase-preserving gain close to $\tilde{\omega}\mathrm{_b}$ (and $\tilde{\omega}\mathrm{_a}$, not shown). The upward frequency shift of the gain curve is due to the Kerr nonlinearity of the hybridized modes (see \cref{A_Gain Fits}). As illustrated in panel (b), applying the conversion pump with power $P\mathrm{_c}$ at frequency $\omega\mathrm{_c}$~=~$\omega\mathrm{_g}$~+~$\tilde{\omega}\mathrm{_a}$~-~$\tilde{\omega}\mathrm{_b}$ in addition to the gain pump, activates beam-splitter interactions between the hybridized modes, resulting in the appearance of a new idler tone and a second peak in the gain profile close to $\tilde{\omega}\mathrm{_b}$. The optimal bandwidth is achieved when the two peaks around $\tilde{\omega}\mathrm{_b}$ coalesce near the BP operational point, as illustrated in panel (c). For all gain curves in (c), the ratio $P\mathrm{_c}$/$P\mathrm{_g}$ remains approximately constant. The black dashed lines in (a) and (c) depict fits obtained with the Bose-Hubbard dimer model (see \cref{A_Gain Fits}). \textbf{(d)}~Comparison of the measured GBW product for single-pumped and double-pumped grAlPA. The green dashed line represents the upper limit, given by the linewidth of the hybridized mode $\kappa\mathrm{_a}$, which can be independently optimized, for example by impedance engineering \cite{Naaman2017,Naaman2022May}. BWs extracted from the fits in (a) and (c) are depicted by the black dashed lines. For the single-pumped grAlPA, the BW scaling is consistent with the measured equivalent damping rate $\kappa\mathrm{_{eq}}/2\pi$~=~19~$\pm$~4~MHz (see \cref{A_circuitParameters}).}
\label{fig_gain_protocol}
\end{figure*}

As shown in~\cref{fig_phase_diagram}(b), operating at both the EP and BP enables signal amplification away from instability, therefore avoiding the GBW limit. 
The BW obtained at the BP approaches the damping rate $\kappa$ (see~\cref{fig_phase_diagram}(c)) and in theory remains completely independent of gain~\cite{Metelmann2022Bog}. 
Moving beyond the BP and away from instability leads to a broader BW. However, as $\mathcal{C}_{\mathrm{BS}}$ increases, the increasing imaginary part of the eigenvalues results in a mode splitting of the output spectrum.
Following this logic, quantum amplifiers operating at the EP and BP are also referred to as gain–conversion (GC) and gain–conversion-imbalance (GCI) amplifiers, respectively~\cite{Chien2020Apr,LannesThesis2020,Jiang2023Apr}.
It should be noted that the choice of pump arrangement is not unique, and alternative configurations could be explored in future studies.


\cref{fig_gain_protocol}(a) shows the resulting power dependent gain profiles close to $\tilde{\omega}\mathrm{_{b}}$ when the grAlPA is driven by a single pump at frequency $\omega\mathrm{_g}/2\pi$~=~8.30172~GHz (see \cref{A_Full gain profiles} for the corresponding gain curves near $\tilde{\omega}\mathrm{_{a}}$). We observe a GBW scaling following the relation GBW~=~$\kappa\mathrm{_{eq}}/G\mathrm{_0}$, as shown in \cref{fig_gain_protocol}(d), where $G\mathrm{_0}$ is the maximum amplifier gain and $\kappa\mathrm{_{eq}}/2\pi$~=~19.2~MHz the calculated equivalent damping rate. We are able to fit all gain profiles simultaneously using the pump line attenuation as the only fitting parameter (see \cref{A_Gain Fits}), which validates the Bose-Hubbard dimer model (cf. \cref{fig_design}(a)).

\begin{figure}[!b]
\includegraphics[width = 1\columnwidth]{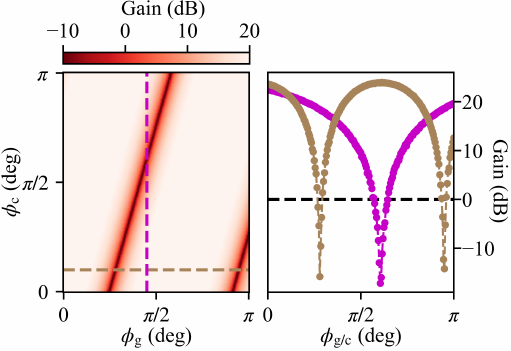}
\caption{\textbf{Phase-dependent gain close to the BP.} The measurements are taken at $\tilde{\omega}\mathrm{_b}$, for a pump configuration giving a maximum gain $G\mathrm{_0}$ in the range of 20~dB. The left panel shows phase-dependent gain as a function of the input phases of the gain pump $\phi\mathrm{_g}$ and conversion pump $\phi\mathrm{_c}$. In the right panel we plot linecuts taken for fixed $\phi\mathrm{_g}$ (purple) and fixed $\phi\mathrm{_c}$ (brown). }
\label{fig_phase_dependent_gain}
\end{figure}

\begin{figure}[!t]
\includegraphics[width = 1\columnwidth]{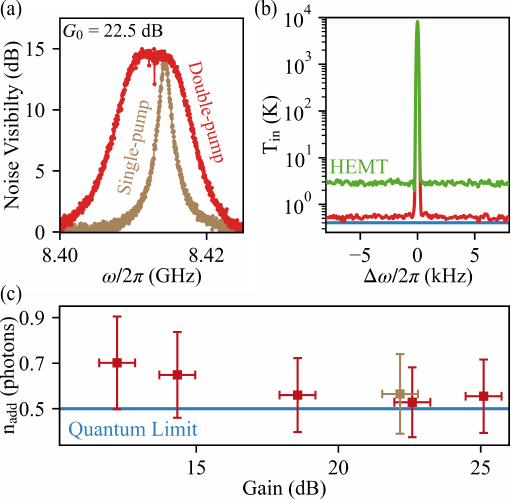}
\caption{\textbf{Noise performance in the phase-preserving regime close to the BP.} \textbf{(a)}~Noise visibility of the grAlPA when driven by a single pump (brown) and two pumps (red), at a maximum gain $G\mathrm{_0}$~=~22.5~dB. \textbf{(b)}~Input-referred noise temperature as a function of detuning $\mathrm{\Delta} \omega$ from a power-calibrated tone at $\omega/2\pi$~=~8.412~GHz. The green and red solid lines correspond to measurements with both pumps off and on, respectively. When both driving tones are on, the noise approaches the standard quantum limit for phase-preserving amplification depicted by the solid blue line. \textbf{(c)}~Added noise of the grAlPA as a function of gain. The quantum limit is defined as half a photon of noise from vacuum fluctuations at the idler frequency. Errorbars represent the uncertainty propagated from the power calibration (see \cref{A_PowerCal}). For comparison, we show the brown point, measured under single-pump operation.}
\label{fig_noise_performance}
\end{figure}

The protocol to achieve an enhanced instantaneous bandwidth with two pumps is shown in \cref{fig_gain_protocol}(b) and the performance for optimally tuned pumps is demonstrated in \cref{fig_gain_protocol}(c).  
By applying a conversion pump at $\omega\mathrm{_c}/2\pi$~=~8.078~GHz in addition to the gain pump $\omega\mathrm{_g}$, we activate beam splitter and single-mode squeezing interactions. Together, these effects give rise to two additional idler tones, which manifest as gain peaks near each dimer mode. The BW is maximized by varying the conversion pump $P\mathrm{_c}$ for fixed gain pump $P\mathrm{_g}$, until the two peaks visible in \cref{fig_gain_protocol}(b) coalesce. This occurs in the vicinity of the BP operating point. For a maximum gain $G\mathrm{_0}$~=~20~dB, we obtain a sixfold bandwidth improvement compared to the single-pump configuration. We repeat the same optimization for different gain pumps and $G\mathrm{_0}$, to obtain the plots shown in \cref{fig_gain_protocol}(c). Remarkably, for $G\mathrm{_0} \sim$~12~dB we obtain a BW surpassing the linewidth of the narrowest dimer mode $\kappa\mathrm{_a}$, as expected when operating near the BP (cf. \cref{fig_phase_diagram}(e)), and we measure a nonconventional GBW limit for up to 25~dB gain (see \cref{fig_gain_protocol}(d)). 

We observe two main discrepancies between the theoretical calculations and the measured gain profiles. First, the obtained BW improvement at 20~dB gain is a factor of two smaller than the expectation from the model. Second, in contrast to the calculated flat-top profile shown in \cref{fig_phase_diagram}(b), a slope is observed at the uppermost region of the optimized gain curve. Both discrepancies arise from the asymmetry in the coupling of the grAl resonators to the unique microwave port in our implementation, a behavior that is nonetheless captured by \cref{eq_linearized_H} when we use the experimental circuit parameters to calculate the gain profiles (see \cref{A_Gain Fits}). This asymmetry can be mitigated in future device designs by adding a second microwave port, which also couples to the right resonator. 

An interesting feature of operating the grAlPA near the BP is the emergence of phase-sensitive amplification at $\tilde{\omega}\mathrm{_a}$ and $\tilde{\omega}\mathrm{_b}$. This effect arises due to the pump frequency configuration, which induces idler degeneracy at $\tilde{\omega}\mathrm{_a}$, subsequently transferred to $\tilde{\omega}\mathrm{_b}$ via the two-mode squeezing interactions of \cref{eq_linearized_H}. 
In \cref{fig_phase_dependent_gain} we show grAlPA phase-dependent gain at $\tilde{\omega}\mathrm{_b}$ near the BP, with a pump configuration that provides close to 20~dB phase-preserving gain. We observe a gain modulation of 41~dB, comparable with values reported in other kinetic inductance materials \cite{Parker2022Mar,Xu2023Feb,Vaartjes2023Nov,Mohamed2023Nov,Frasca2023Dec}. The strong -17~dB deamplification level recommends our amplifiers as efficient sources for single-mode vacuum squeezing for future quantum optics experiments.

A crucial question to address is whether the noise performance of the amplifier is degraded by the double-pump configuration. 
We evaluate the amplifier noise visibility, defined as the excess noise power of the readout line compared to when the grAlPA is off, and find that operating the grAlPA at 22.5~dB gain yields a noise level unaffected by activation of the conversion pump, as shown in Fig.~\ref{fig_noise_performance}(a).
Moreover, by using a power calibrated (see \cref{A_PowerCal}) pilot tone detuned 5~MHz above $\tilde{\omega}\mathrm{_b}$, we determine the amplifier input-referred noise level when operated in phase-preserving mode. As presented in \cref{fig_noise_performance}(b) when the pumps are off, the noise floor is given by a High Electron Mobility Transistor (HEMT) amplifier, resulting in a noise temperature of about 3~K, consistent with the manufacturer's datasheet \cite{HEMT_LNF} and the expected insertion loss of the microwave connections. By double-pumping the grAlPA at 22.5~dB gain, the input-referred noise approaches the standard quantum limit for phase-preserving amplifiers \cite{Caves1982Oct}, indicated by the blue line in \cref{fig_noise_performance}(b). We calculate the total added photon noise $n\mathrm{_{add}}=k\mathrm{_B}\overline{T}\mathrm{_{in}}$/$\hbar\omega\mathrm{_s}-n\mathrm{_{s}}$, where $\overline{T}\mathrm{_{in}}$ is the noise floor extracted from the power spectral density (cf.~\cref{fig_noise_performance}(b)) with the amplifier on and $n\mathrm{_{s}}$~=~0.5 is the noise accompanying the pilot tone. 
As shown in \cref{fig_noise_performance}(c), the initial decrease in $n\mathrm{_{add}}$ with increasing gain results from saturation of noise in the amplification stages after the grAlPA, while beyond 18~dB, the noise level approaches the quantum limit within less than half-photon uncertainty.

In summary, we have demonstrated a standing-wave parametric amplifier with a nonconventional gain–bandwidth scaling, realized by double-pumping a granular aluminum Bose–Hubbard dimer. 
Unlike previous implementations, our approach relies solely on four-wave mixing stemming from grAl Kerr nonlinearity, offering greater integration flexibility by eliminating the need to engineer pure three-wave mixing processes in the device.
The device achieves a factor of six bandwidth broadening for a maximum gain of 20~dB, with improved GBW scaling up to 25 dB, and it maintains added noise levels near the quantum limit. The large amplification–deamplification ratio observed in phase-sensitive mode, also highlights the promise of these amplifiers for generating both single and two-mode vacuum squeezing. These features could be investigated more efficiently by designing a two-port device with direct access to the hybridized quadratures, which would allow us to probe the path entanglement of the emitted light. Furthermore, the concept of a multi-pumped parametric amplifier could be extended by adding a third mode and applying six pump tones, which has been predicted to also provide nonreciprocity~\cite{Metelmann2015Jun,Metelmann2022Bog}.

All relevant data are available from the corresponding
author upon reasonable request.

\begin{acknowledgments}
We are grateful to Gary Steele for fruitful discussions and we acknowledge technical support from S. Diewald and L. Radtke. 
Funding was provided by the Deutsche Forschungsgemeinschaft (DFG – German Research Foundation) under project number 450396347 (GeHoldeQED), the European Union under the Horizon Europe Program, grant agreement number 101080152 (TruePA) and the Federal Ministry of Research, Technology and Space (BMBFTR) within project QSolid (FKZ:13N16151). Facilities use was supported by the Karlsruhe Nano Micro Facility (KNMFi) and KIT Nanostructure Service Laboratory (NSL). We acknowledge qKit for providing a convenient measurement software framework.
\end{acknowledgments}

\appendix

\section*{Supplemental material}

\section{Theory of multi-pumped Bose-Hubbard dimer}
\label{A_Theory}

\subsection{Bare basis}
\label{A_Double_pump_bare_basis}

The dynamics of the Bose-Hubbard dimer illustrated in~\cref{fig_design}(a), can be expressed by the following Hamiltonian
\begin{subequations}
\begin{align}
    \label{H123}
    \hat H  & =  \hat H\s{BL} + \hat H\s{Kerr},
    \\
    \text{\vspace{1.5em}} \nonumber \\
    \hat H\s{BL} / \hbar & = \omega\s{L} \: \hat a\s{L}^{\dagger} \hat a\s{L} + \omega\s{R} \: \hat a\s{R}^{\dagger} \hat a\s{R} + J \left(\hat a\s{L}^{\dagger} \hat a\s{R} + \hat a\s{R}^{\dagger} \hat a\s{L} \right),
    \\
    \text{\vspace{1.5em}} \nonumber \\
    \hat H\s{Kerr} / \hbar  & =  \frac{\mathcal{K}\s{L}}{2} \: \hat a\s{L}^{\dagger} \hat a\s{L}^{\dagger} \hat a\s{L} \hat a\s{L} + \frac{\mathcal{K}\s{R}}{2} \: \hat a\s{R}^{\dagger} \hat a\s{R}^{\dagger} \hat a\s{R} \hat a\s{R},
\end{align}
\end{subequations}
where $\hat H\s{BL}$ and $\hat H\s{Kerr}$ describe the bilinear and nonlinear parts of the Hamiltonian, respectively. Here, $\omega_j$ (with $j = \mathrm{L} , \mathrm{R}$) denotes the resonance frequency of the $j$ resonator, $J$ represents their hopping interaction strength, and $\mathcal{K}_j$ ($\mathcal{K}_j<0$) denotes their Kerr nonlinearity. 

When the system is strongly driven, the dynamics can be linearized by decomposing each mode operator into a classical mean-field amplitude and a fluctuation operator,
\begin{align}
\label{Mean_Bare}
\hat{a}\s{L/R} & = \alpha\s{L/R}  +  \delta  \hat a\s{L/R}, 
\end{align}
where, $\alpha\s{L}$ and $\alpha\s{R}$ denote the mean-field components of resonators L and R, respectively. In the presence of two pumps with frequencies $\omega\s{g}$ and $\omega\s{c}$, the mean-field amplitudes can be written as 

\begin{align}
\label{Ansatz_two}
\alpha_j (t) = \alpha\mathrm{^g_\textit{j}} \, \mathrm{e}^{- i \omega\mathrm{_g} t} + \alpha\mathrm{^c_\textit{j}} \, \mathrm{e}^{- i \omega\mathrm{_c} t},
\end{align}
where $j = \mathrm{L, R}$.

The linearized bare Hamiltonian of the system is then expressed as
\begin{align}
\label{HB_lin}
\hat H\s{B} (t) / \hbar & = \tilde{\omega}\s{L}(t) \, \hat{a}\s{L}^{\dag} \hat{a}\s{L} + \tilde{\omega}\s{R} (t) \, \hat{a}\s{R}^{\dag} \hat{a}\s{R} + J \left(\hat a\s{L}^{\dag} \hat a\s{R}  + \mathrm{h.c.} \right) \nonumber 
\\
\\
& + \left( \tilde{\mathcal{K}}\s{L} (t) \, \hat{a}\s{L}^{\dag} \hat{a}\s{L}^{\dag} + \mathrm{h.c.} \right) + \left( \tilde{\mathcal{K}}\s{R} (t) \, \hat{a}\s{R}^{\dag} \hat{a}\s{R}^{\dag} + \mathrm{h.c.} \right), \nonumber 
\end{align}
where, for simplicity, we have redefined $\delta \hat{a}\s{L/R} \equiv \hat{a}\s{L/R}$. The general Kerr-induced frequency shifts of the bare resonators are given by
\begin{align}
\label{Par_Wshift}
\tilde{\omega}_j(t) & = \omega_j + 2 \mathcal{K}_j |\alpha_j(t)|^2  = \omega_j +  2 \mathcal{K}_j \left( |\alpha\mathrm{^{g}_\textit{j}}|^2 + \alpha\mathrm{^{c}_\textit{j}}|^2 \right) \nonumber \\
&  + 2 \, \mathcal{K}_j \left( \alpha\mathrm{^{g}_\textit{j}} \alpha\mathrm{^{c}_\textit{j}}^* \, \mathrm{e}^{-i \Delta_p t} + \mathrm{h.c.} \right), 
\end{align}
where $\Delta_p = \omega\s{g} - \omega\s{c}$. Moreover, the effective time-dependent Kerr coefficients take the form 
\begin{align}
\label{Par_Kerrshift}
\tilde{\mathcal{K}}_j (t) & = \frac{\mathcal{K}_j}{2} \alpha_j^2(t)  = \frac{\mathcal{K}_j}{2} \Big( (\alpha\mathrm{^{g}_\textit{j}})^2 \, \mathrm{e}^{- 2 i \omega_{g} t} + (\alpha\mathrm{^{c}_\textit{j}})^2 \, \mathrm{e}^{- 2 i \omega_{c} t}  \nonumber \\
& + 2  \, \alpha\mathrm{^{c}_\textit{j}} \alpha\mathrm{^{g}_\textit{j}} \, \mathrm{e}^{-i (\omega_{g} + \omega_{c}) t} \Big).
\end{align}

Transforming each mode into an appropriate rotating frame defined by the reference frequency $\tilde{\omega}\s{a} = (\omega\s{g} + \omega\s{c})/2$ such that $\hat{a}\s{L} \rightarrow \hat{a}\s{L} \mathrm{e}^{- i \tilde{\omega}\s{a} t}$, and $\hat{b}\s{R} \rightarrow \hat{b}\s{R} \mathrm{e}^{- i \tilde{\omega}\s{a} t}$,~\cref{HB_lin} is then is given by
\begin{align}
    \label{HB_lin_Rxy}
    \hat H\s{B} / \hbar & = \big( \tilde{\omega}\s{L} - \tilde{\omega}\s{a} + \mathcal{K}\s{1,L} \, \mathrm{e}^{-i \Delta_p t} + \mathcal{K}\s{2,L} \, \mathrm{e}^{i \Delta_p t} \Big) \, \hat a\s{L}^{\dagger} \hat a\s{L} \nonumber  \\
    & +  \Big( \tilde{\omega}\s{R} - \tilde{\omega}\s{a} + \mathcal{K}\s{1,R} \, \mathrm{e}^{-i \Delta_p t} + \mathcal{K}\s{2,R} \, \mathrm{e}^{i \Delta_p t} \Big) \, \hat a\s{R}^{\dagger} \hat a\s{R} \nonumber \\ 
    & + \frac{1}{2} \Big[ \Big( \mathcal{K}\s{3,L} \, \mathrm{e}^{- i \Delta_p  t} + \mathcal{K}\s{4,L} \, \mathrm{e}^{ i \Delta_p  t} \nonumber + \mathcal{K}\s{5,L} \Big) \hat a\s{L}^{\dagger}  a\s{L}^{\dagger}  + \mathrm{h.c.} \Big] \nonumber \\  
   & + \frac{1}{2} \Big[ \Big( \mathcal{K}\s{3,R} \, \mathrm{e}^{- i \Delta_p t} + \mathcal{K}\s{4,R} \, \mathrm{e}^{ i \Delta_p t} \nonumber + \mathcal{K}\s{5,R} \Big) \hat a\s{R}^{\dagger}  a\s{R}^{\dagger}  + \mathrm{h.c.} \Big] \nonumber,\\
\end{align}
where, the effective interaction parameters are defined as

\begin{subequations}
\label{KK_eff}
\begin{align}
\mathcal{K}\s{1,j} & = 2 \mathcal{K}\s{j} \, \alpha\mathrm{^{g}_\textit{j}}  \alpha\mathrm{^{c}_\textit{j}}^*, \\
\mathcal{K}\s{2,j} & = \mathcal{K}\s{1,j}^* = 2 \mathcal{K}\s{j} \, \alpha\mathrm{^{g}_\textit{j}}^*  \alpha\mathrm{^{c}_\textit{j}}, \\
\mathcal{K}\s{3,j} & =  \mathcal{K}\s{j} \, (\alpha\mathrm{^{g}_\textit{j}})^2, \\
\mathcal{K}\s{4,j} & =  \mathcal{K}\s{j} \, (\alpha\mathrm{^{c}_\textit{j}})^2,\\
\mathcal{K}\s{5,j} & =  2 \mathcal{K}\s{j} \, \alpha\mathrm{^{g}_\textit{j}}  \alpha\mathrm{^{c}_\textit{j}}.
\end{align}
\end{subequations}
The dynamics of the fluctuations can be determined by the Heisenberg-Langevin equations 
\begin{subequations}
\label{LE_tot}
\begin{align}
\frac{d \hat a\s{L}}{d t} & = i \left( \Delta\s{L} + i \frac{\kappa}{2} \right) \, \hat a\s{L} 
- i J a\s{R}   - i \mathcal{K}\s{5,L} \, \hat{a}\s{L}^\dagger  \label{KK_eff1 }\\
& - i \mathcal{K}\s{1,L} \, \mathrm{e}^{-i \Delta_p t} \, \hat a\s{L} - i \mathcal{K}\s{2,L} \, \mathrm{e}^{+i \Delta_p t} \, \hat a\s{L} \nonumber \\
&- i \mathcal{K}\s{3,L} \, \mathrm{e}^{-i \Delta_p t}  \, \hat{a}\s{L}^\dagger 
 - i \mathcal{K}\s{4,L} \,\mathrm{e}^{i \Delta_p t}  \, \hat{a}\s{L}^\dagger - \sqrt{\kappa} \, \hat{a}\s{L,in}, \nonumber \\
\frac{d \hat a\s{R}}{d t} & = i \Delta\s{R} \, \hat a\s{R} - i Ja\s{L}  - i \mathcal{K}\s{5,R} \, \hat{a}\s{R}^\dagger  \label{KK_eff2 }\\
& - i \mathcal{K}\s{1,R} \, \mathrm{e}^{-i \Delta_p t} \, \hat a\s{R} - i \mathcal{K}\s{2,R} \, \mathrm{e}^{+i \Delta_p t} \, \hat a\s{R} \nonumber \\ 
& - i \mathcal{K}\s{3,R} \, \mathrm{e}^{-i \Delta_p t}  \, \hat{a}\s{R}^\dagger
 - i \mathcal{K}\s{4,R} \,\mathrm{e}^{i \Delta_p t}  \, \hat{a}\s{R}^\dagger, \nonumber
\end{align}
\end{subequations}
where $\Delta_{j} =  (\omega\s{g} + \omega\s{c})/2 - \tilde{\omega}_j$. Alongside the fluctuation dynamics described above, the mean-field solutions are determined by solving the following set of equations
\begin{subequations}
\label{MF_Bare}
\begin{align}
 \alpha\mathrm{^\textit{v}_R} 
 \delta\s\mathrm{_{\textit{v},R}} 
 & = J \alpha\mathrm{^\textit{v}_L}, \\
 \alpha\mathrm{^\textit{v}_L} \left( \delta\s\mathrm{_{\textit{v},L}} + i \frac{\kappa}{2} \right) & = J \alpha\mathrm{^\textit{v}_R} - i \sqrt{\kappa} \, \alpha\s{v,in},
\end{align}
\label{eq_steady_state_Langevin_alphas}
\end{subequations}
where $v = \mathrm{g,c}$ and 
\begin{subequations}
\label{MF_Bare_detun}
\begin{align}
\delta\s{g,L} & = \omega\s{g} - \omega\s{L} - \mathcal{K}\s{L} \left(|\alpha\mathrm{^{g}_L}|^2 + 2 |\alpha\mathrm{^{c}_L}|^2 \right), \\
\delta\s{c,L} & = \omega\s{c} - \omega\s{L} - \mathcal{K}\s{L} \left( 2 |\alpha\mathrm{^{g}_L}|^2 + 2 |\alpha\mathrm{^{c}_L}|^2 \right),\\
\delta\s{g,R} & = \omega\s{g} - \omega\s{R} - \mathcal{K}\s{R} \left(|\alpha\mathrm{^{g}_R}|^2 + 2 |\alpha\mathrm{^{c}_R}|^2 \right), \\
\delta\s{c,R} & = \omega\s{c} - \omega\s{R} - \mathcal{K}\s{R} \left(2 |\alpha\mathrm{^{g}_R}|^2 + |\alpha\mathrm{^{c}_R}|^2 \right),  \\
\end{align}
\end{subequations}
where the input amplitudes are defined as $|\alpha\s{g,in}|$~=~$\sqrt{P\s{g}/(\hbar \omega\s{g})}$, $|\alpha\s{c,in}|$~=~$\sqrt{P\s{c}/(\hbar \omega\s{c})}$.
It should be noted that the selection of this particular rotating frame is solely for computational convenience. As all terms in the system are included without applying the rotating-wave approximation (RWA), the chosen frame of rotation has no impact on the physical results.

By taking the Fourier transform of \cref{LE_tot}, and using the input-output relation $\hat{a}\mathrm{_{L,out}}=\sqrt{\kappa}\,\hat{a}_\mathrm{L}+\hat{a}\mathrm{_{L,in}}$, we obtain the scattering matrix of the bare system, which in this case is given by 

\begin{equation}
\label{SM_B}
    \mathbf{S}(\omega) = \mathbf{K}\s{\mathrm{ext}} \: \mathbf{M}^{-1}(\omega) \: \mathbf{K} + \mathbf{1}_{20},
\end{equation}
where
\begin{align}
\label{Kext}
\mathbf{K}\s{\mathrm{ext}} = \begin{pmatrix}
\sqrt{\kappa} \, \mathbf{1}_{10} & O \\
O & O
\end{pmatrix}_{20 * 20}.
\end{align}
and
\begin{align}
\label{KB}
\mathbf{K} = \begin{pmatrix}
\sqrt{\kappa} \, \mathbf{1}_{10} & O \\
O & \sqrt{\kappa\s{R}} \, \mathbf{1}_{10}
\end{pmatrix}_{20 * 20}.
\end{align}
\subsubsection{Dynamics in presence of a single pump}
\label{A_Theory_Single_Pump_bare}

If only the gain pump is applied, then $\alpha\mathrm{^{c}_{L/R}}=0$ and the mean-field Langevin equations \cref{eq_steady_state_Langevin_alphas} reduce to 

\begin{subequations}
\label{MF_Bare_single}
\begin{align}
\alpha_\mathrm{R}  \delta\s{L} & = J \alpha\s{L}, \\
 \alpha\s{L} \left( \delta\s{R}  + i \frac{\kappa}{2} \right) & = J \alpha\mathrm{_R} - i \sqrt{\kappa} \, \alpha\s{g,in}.
\end{align}
\label{eq_langevin_bare_basis}
\end{subequations}

with $\delta_j=\omega\s{g}-\omega_j-\mathcal{K}_j\, |\alpha\mathrm{_{j}}|^2$.

Moreover, the equation of motion for the fluctuations simplifies to 

\begin{subequations}
\label{LELB}
    \begin{align}
     \frac{d \hat a\s{L}}{d t} & =  i \left( \Delta\s{L}  + i \frac{\kappa}{2} \right) \hat{a}\s{L} - i J \hat a\s{R} - i \tilde{\mathcal{K}}\s{L} \hat a\s{L}^{\dagger} - \sqrt{\kappa} \hat a\s{L,in}, \\
     \frac{d \hat a\s{R}}{d t} & = i \left( \Delta\s{R} + \frac{\kappa\s{R}}{2} \right) \hat{a}\s{R} - i J \hat a\s{L} - i \tilde{\mathcal{K}}\s{R} \hat a\s{R}^{\dagger},
\end{align}
\end{subequations}
where
\begin{subequations}
\begin{align}
\label{delta_single}
\tilde{\mathcal{K}}_j & = \frac{\mathcal{K}_j}{2} \alpha_j^2.  \\
\Delta_{j} & = \omega\s{g} -\omega_j -  2 \mathcal{K}_j |\alpha_{j}|^2,  
\end{align}
\end{subequations}
\subsection{Hybridized basis}
\label{A_Double_pump_hybrid_basis}

Due to the hopping interaction $J$, the system energy levels hybridize. The bilinear Hamiltonian $\hat H_{\mathrm{BL}}$ can be diagonalized by introducing hybridized mode operators defined as
 \begin{align}  
     \label{AB}
     \hat c\s{a} & = - \sin \theta \: \hat a\s{L} + \cos \theta \: \hat a\s{R},\nonumber 
    \\
    \\
    \hat c\s{b} & = \cos \theta \: \hat a\s{L} + \sin \theta \: \hat a\s{R}, \nonumber  
\end{align}
with corresponding mode frequencies
\begin{equation}
    \label{eq_hybridized_frquencies}
    \omega\s{a/b}  = \omega_{+} \mp \sqrt{J^2 + \omega_-^2}, 
\end{equation}
\noindent
where $\omega_{\pm} = (\omega\mathrm{_L} \pm \omega\mathrm{_R}) /2$, and $\tan 2 \theta = J / \omega_-$ ( $\omega\mathrm{_R} \geq \omega\mathrm{_L} $). 
In the hybridized basis, the full Hamiltonian takes the form
\begin{align}
\label{HHyb}
    \hat H / \hbar  & = \sum_{i=a,b} \omega_i \, \hat c_i^{\dagger} \hat c_i +\frac{\mathcal{K}\s{aa}}{2} \, \hat c\s{a}^{\dagger} \hat c\s{a}^{\dagger} \hat c\s{a}  \hat c\s{a}  + \frac{\mathcal{K}\s{bb}}{2} \, \hat c\s{b}^{\dagger} \hat c\s{b}^{\dagger} \hat c\s{b} \hat c\s{b} \nonumber \\
    & + \frac{\mathcal{K}\s{ab}}{8} \left( \hat c\s{a}^{\dagger} \hat c\s{a}^{\dagger} \hat c\s{b} \hat c\s{b} + h.c \right) + \frac{\mathcal{K}\s{ab}}{2} \hat c\s{a}^{\dagger} \hat c\s{a} \hat c\s{b}^{\dagger} \hat c\s{b}   \\
   & - \frac{\mathcal{K}\s{ab}}{4} \left[ \mu_- \left( \hat c\s{a}^{\dagger} \hat c\s{a}^{\dagger} \hat c\s{a} \hat c\s{b} + \mathrm{h.c.} \right) + \mu_+ \left( \hat c\s{b}^{\dagger} \hat c\s{b}^{\dagger} \hat c\s{b}  \hat c\s{a}  + \mathrm{h.c.} \right) \right]. \nonumber
\end{align}
where $\omega_{i}$ ($i = \mathrm{a, b}$) are the effective hybridized mode frequencies. The coefficients $\mathcal{K}\s{aa}$, $\mathcal{K}\s{bb}$, and $\mathcal{K}\s{ab}$ denote the collective self-Kerr and cross-Kerr nonliearities and are given by
\begin{subequations}
\begin{align}
\label{Kerr}
    \mathcal{K}\s{ab} & = \frac{J^2 \left( \mathcal{K}\s{L} + \mathcal{K}\s{R} \right)}{J^2 + \omega_-^2}, \\
    \mathcal{K}\s{aa} & = \frac{1}{4} \mathcal{K}\s{ab} \left( 1 + \frac{2 \omega_-^2}{J^2} \right) - \frac{\left( \mathcal{K}\s{L} - \mathcal{K}\s{R} \right) \: \omega_-}{2 \sqrt{J^2 + \omega_-^2}}, \\
    \mathcal{K}\s{bb} & = \frac{1}{4} \mathcal{K}\s{ab} \left( 1 + \frac{2 \omega_-^2}{J^2} \right) + \frac{\left( \mathcal{K}\s{L} - \mathcal{K}\s{R} \right) \: \omega_-}{2 \sqrt{J^2 + \omega_-^2}}.
\end{align}
\end{subequations}

In~\cref{HHyb}, the term $\frac{\mathcal{K}\s{ab}}{2} \hat c\s{a}^{\dagger} \hat c\s{a} \hat c\s{b}^{\dagger} \hat c\s{b}$ corresponds to cross-Kerr interactions and $\tfrac{\mathcal{K}\s{ab}}{8} \left( \hat c\s{a}^{\dagger} \hat c\s{a}^{\dagger} \hat c\s{b} \hat c\s{b} + \text{h.c.} \right)$ describes two-photon exchange processes. The last two terms in Hamiltonian~\eqref{HHyb} can be regarded as higher-order corrections. Additionally, the coefficients $\mu_{\mp}$, which quantify the frequency shift induced by the nonlinear hopping interaction ($ \propto (\hat c\s{a/b}^{\dagger} \hat c\s{a/b}^{\dagger} \hat c\s{a/b} \hat c\s{b/a} + \mathrm{h.c.} $)), are given by
\begin{equation}
\label{mu}
     \mu_{\mp} = \sqrt{1 + \frac{\omega_-^2}{J^2}} \: \frac{\left( \mathcal{K}\s{L} - \mathcal{K}\s{R} \right)}{\left( \mathcal{K}\s{L} + \mathcal{K}\s{R} \right)} \mp \frac{\omega_-}{J}.
\end{equation}
%


When two pumps are applied, the Hamiltonian of the system in the hybridized basis is given by
\begin{align}
   \label{H_lin_H}
    \hat H\s{H} (t) / \hbar & = \sum_{i = \mathrm{a}, \mathrm{b}} \Big \lbrace \tilde{\omega}_i (t) \: \hat c_i^{\dagger} \hat c_i  + \Big [\Big( \Lambda\mathrm{_{S_\textit{i}}}^{(1)} \: \mathrm{e}^{-2 i \omega\s{g} t} + \Lambda\mathrm{_{S_\textit{i}}}^{(2)} \: \mathrm{e}^{-2 i \omega\s{c} t} \nonumber 
    \\
    & + \Lambda\mathrm{_{S_\textit{i}}}^{(12)} \: \mathrm{e}^{- i (\omega\s{g} + \omega\s{c}) t} \Big) \: \hat c_i^{\dagger} \hat c_i^{\dagger} + \mathrm{h.c} \Big] \Big \rbrace  +  \Big[ \Big( \Lambda\s{TMS}^{(1)} \: \mathrm{e}^{-2 i \omega\s{g} t} \nonumber  \\
    & +\Lambda\s{TMS}^{(2)} \: \mathrm{e}^{-2 i \omega\s{c} t} + \Lambda\s{TMS}^{(12)} \: \mathrm{e}^{- i (\omega\s{g} + \omega\s{c}) t} \Big) \: \hat c_a^{\dagger} \hat c_b^{\dagger} + \mathrm{h.c} \Big] \nonumber \\
    & + \Big[ \Lambda\s{BS}^{(0)} + \left( \Lambda\s{BS}^{(12)} \: \mathrm{e}^{-i (\omega\s{g} - \omega\s{c}) t} + \mathrm{h.c} \right) \Big] \left(\hat c_a^{\dagger} \hat c_b + \hat c_b^{\dagger} \hat c_a \right).
\end{align}
where $\tilde{\omega}\mathrm{_{a/b}}$ are the Kerr-shifted frequencies given by
\begin{align}
\label{WAB_HL}
\tilde{\omega}\s{a}  & = \omega\s{a}  + 2 \mathcal{K}\s{L} \left[ |\alpha\mathrm{^{g}_{L}}|^2 + |\alpha\mathrm{^{c}_{L}}|^2 + 2 \left( \alpha\mathrm{^{g}_{L}} \alpha\mathrm{^{c}_{L}}^* \, \mathrm{e}^{-i \Delta_p t} + \mathrm{h.c.} \right) \right] \sin^2 \theta  \nonumber \\
& + 2 \mathcal{K}\s{R} \left[|\alpha\mathrm{^{g}_{R}}|^2 + |\alpha\mathrm{^{c}_{R}}|^2 + 2 \left( \alpha\mathrm{^{g}_{R}} \alpha\mathrm{^{c}_{R}}^* \, \mathrm{e}^{-i \Delta_p t} + \mathrm{h.c.} \right)  \right] \cos^2\theta , \nonumber \\
\\
\tilde{\omega}\s{b}   & = \omega\s{b}  +  2 \mathcal{K}\s{L} \left[ |\alpha\mathrm{^{g}_{L}}|^2 + |\alpha\mathrm{^{c}_{L}}|^2 + 2 \left( \alpha\mathrm{^{g}_{L}} \alpha\mathrm{^{c}_{L}}^* \, \mathrm{e}^{-i \Delta_p t} + \mathrm{h.c.} \right) \right] \cos^2 \theta  \nonumber \\
& + 2 \mathcal{K}\s{R} \left[|\alpha\mathrm{^{g}_{R}}|^2 + |\alpha\mathrm{^{c}_{R}}|^2 + 2 \left( \alpha\mathrm{^{g}_{L}} \alpha\mathrm{^{c}_{L}}^* \, \mathrm{e}^{-i \Delta_p t} + \mathrm{h.c.} \right) \right] \sin^2\theta  , \nonumber
\end{align}
and the single-mode squeezing couplings are defined as
\begin{align}
\label{LSH}
\Lambda\s{S\s{a}}^{(1)} & =\mathcal{K}\s{L} (\alpha\mathrm{^{g}_{L}})^2 \sin^2 \theta + \mathcal{K}\s{R} (\alpha\mathrm{^{g}_{R}})^2 \cos^2 \theta, \nonumber \\
\Lambda\s{S\s{a}}^{(2)} & = \mathcal{K}\s{L} (\alpha\mathrm{^{c}_{L}})^2 \sin^2 \theta + \mathcal{K}\s{R} (\alpha\mathrm{^{c}_{R}})^2 \cos^2 \theta, \\
\Lambda\s{S\s{a}}^{(12)} & = 2 \left( \mathcal{K}\s{L} \alpha\mathrm{^{g}_{L}} \alpha\mathrm{^{c}_{L}} \sin^2 \theta + \mathcal{K}\s{R} \alpha\mathrm{^{g}_{R}} \alpha\mathrm{^{c}_{R}} \cos^2 \theta \right). \nonumber 
\end{align}
Analogous expressions hold for $\Lambda\s{S\s{b}}$, with $\cos^2 \theta$ and $\sin^2 \theta$ interchanged.
In addition, the two-mode squeezing and beam-splitter couplings are, respectively, given by
\begin{align}
\label{LPAH}
\Lambda\s{TMS}^{(1)} & = -\frac{1}{2} \left( \mathcal{K}\s{L} (\alpha\mathrm{^{g}_{L}})^2 - \mathcal{K}\s{R} (\alpha\mathrm{^{g}_{R}})^2 \right) \sin 2\theta,  \nonumber \\
\Lambda\s{TMS}^{(2)} & = -\frac{1}{2} \left( \mathcal{K}\s{L} (\alpha\mathrm{^{c}_{L}})^2 - \mathcal{K}\s{R} (\alpha\mathrm{^{c}_{R}})^2 \right) \sin 2\theta, \\
\Lambda\s{TMS}^{(12)} & = -\left( \mathcal{K}\s{L} \alpha\mathrm{^{g}_{L}} \alpha\mathrm{^{c}_{L}} - \mathcal{K}\s{R} \alpha\mathrm{^{g}_{R}} \alpha\mathrm{^{c}_{R}} \right) \sin 2\theta, \nonumber 
\end{align}
\begin{align}
\label{LFCH}
\Lambda\s{BS}^{(0)} & = -\Big[ \mathcal{K}\s{L} \left(|\alpha\mathrm{^{g}_{L}}|^2 + |\alpha\mathrm{^{c}_{L}}|^2 \right) - \mathcal{K}\s{R} \left(|\alpha\mathrm{^{g}_{R}}|^2 + |\alpha\mathrm{^{c}_{R}}|^2 \right) \Big] \sin 2\theta, \nonumber \\
\\
\Lambda\s{BS}^{(12)} & = -\left( \mathcal{K}\s{L} \alpha\mathrm{^{g}_{L}}^* \alpha\mathrm{^{c}_{L}} - \mathcal{K}\s{R} \alpha\mathrm{^{g}_{R}}^* \alpha\mathrm{^{c}_{R}} \right) \sin 2\theta. \nonumber
\end{align}
As inferred from~\cref{H_lin_H}, by choosing an appropriate rotating frame with frequency $\omega\s{o}$, certain interaction terms become dominant in the system dynamics, while rapidly oscillating contributions can be neglected. Under this condition, the Hamiltonian~\eqref{H_lin_H} can be rewritten into a time-independent form following
\begin{align}
   \label{H_lin_HM}
   \hat{H}\s{H}/\hbar  & = -\sum_{i= \mathrm{a,b}} \left \lbrace \Delta_i \,\hat c^{\dagger}_{i} \hat c_{i} - \left(\Lambda_{\mathrm{S}_{i}} \, \hat c^{\dagger}_{i} \hat c^{\dagger}_{i} + \mathrm{h.c.} \right) \right \rbrace \\
 & + \Big( \Lambda\mathrm{_{TMS}} \, \hat c^{\dagger}_{\mathrm{a}} \hat c^{\dagger}_{\mathrm{b}} + \mathrm{h.c.} \Big) +
\Big( \Lambda\mathrm{_{BS}} \,\hat c^{\dagger}_{\mathrm{a}} \hat c_{\mathrm{b}} + \mathrm{h.c.} \Big), \nonumber 
\end{align}
where $\Delta_i = \omega\s{o} - \tilde{\omega}\s{i}$. Rotating the Hamiltonian with frequency $\omega\s{o} = (\omega\s{g} + \omega\s{c}) /2$, and after ignoring the fast rotating terms, the main coupling interactions are governed by the set of $\lbrace  \Lambda\s{S\s{a}}^{(12)}, \Lambda\s{S\s{b}}^{(12)}, \Lambda\s{TMS}^{(12)}, \Lambda\s{BS}^{(0)} \rbrace$. On the other hand, rotating the Hamiltonian with frequency $\omega\s{o} = \omega\s{g}$, the main coupling interactions are now $\lbrace  \Lambda\s{S\s{a}}^{(1)}, \Lambda\s{S\s{b}}^{(1)}, \Lambda\s{TMS}^{(1)}, \Lambda\s{BS}^{(0)} \rbrace$.

\subsubsection{Symmetric coupling of bare grAl resonators}
\label{A_Eigenvalues}
To obtain a better understanding of the physical properties of the system, we represent the dynamics as a function of its quadrature operators. The Hamiltonian in the hybridized basis can be written as
\begin{align}
   \label{HXP}
   \hat{H}\s{H}/\hbar  & = -\sum_{i= \mathrm{a,b}} \Big \lbrace \left( \frac{\Delta_i}{2} + \Lambda_{\mathrm{S}_{i}} \right) \hat{X}_i^2  + \left( \frac{\Delta_i}{2} - \Lambda_{\mathrm{S}_{i}} \right) \hat{P}_i^2 \Big \rbrace \nonumber \\
   \\
   & + \left( \Lambda\mathrm{_{BS}} + \Lambda\mathrm{_{TMS}} \right) \hat{X}\s{a} \hat{X}\s{b} + \left( \Lambda\mathrm{_{BS}} - \Lambda\mathrm{_{TMS}} \right) \hat{P}\s{a} \hat{P}\s{b}, \nonumber 
\end{align}
where $\hat X_j = (\hat c_j^{\dagger} + \hat c_j)/ \sqrt{2}$, $\hat P_i = i (\hat c_i^{\dagger} - \hat c_i)/ \sqrt{2}$ ($i = \mathrm{a}, \mathrm{b}$). 
In this representation, the set of equations of motion can be written as 
\begin{equation}
\label{QLE_XP}
    \frac{d \vec{\hat{X}}}{dt} = \mathbf{M}\s{xp} \, \vec{\hat X} + \mathbf{K}\s{xp} \, \vec{\hat X}\s{in},
\end{equation}
where  $\vec{\hat{X}} = [ \hat X_a, \hat P_a, \hat X_b, \hat P_b]^T$ is the quadrature vector, and  $\vec{\hat{X}}\s{in} = [\hat X\s{a,in}, \hat P\s{a,in}, \hat X\s{b,in}, \hat P\s{b,in}]^T$ denotes the quadratures of the input excitations. Additionally, the drift matrix $\mathbf{M}\s{xp}$ and the damping matrix $\mathbf{K}\s{xp}$ are defined as
\begin{equation}
    \label{MXP}
\mathbf{M}\s{xp} =  \scalebox{0.8}{$\begin{pmatrix}
         -\kappa/2 & -\left(\Delta\s{a}+2 \Lambda\s{S\s{a}}\right) & 0 & \Lambda\s{TMS}-\Lambda\s{BS} \\
         \Delta\s{a}-2 \Lambda\s{S\s{a}} & -\kappa/2 & \Lambda\s{TMS}+\Lambda\s{BS} & 0 \\
        0 & \Lambda\s{TMS}-\Lambda\s{BS} & -\kappa/2 & -\left(\Delta\s{b} + 2 \Lambda\s{S\s{b}} \right) \\
         \Lambda\s{TMS}+\Lambda\s{BS} & 0 & \Delta\s{b}-2 \Lambda\s{S\s{b}} &  -\kappa/2
\end{pmatrix}$},
\end{equation}
\begin{align}
\label{KXP}
\mathbf{K}\s{xp} = \sqrt{k} \, \mathbf{1}\s{\mathrm{4}}. 
\end{align}
In order to simplify the drift matrix $\mathbf{M}\s{xp}$ and the damping matrix $\mathbf{K}\s{xp}$, we assumed a symmetric configuration, where the two resonators in~\cref{fig_design}(a) are coupled to two different microwave ports with identical rates $\kappa$. Under this assumption, the linewidths of the hybridized modes reduce to $\kappa\s{a} = \kappa\s{b} = \kappa$. Notice that for the design of~\cref{fig_design}(a), $\kappa_\mathrm{a/b}$ follow instead~\cref{eq_kappas_omegas_hybrid}, which for perfectly hybridized resonators, we obtain $\kappa\s{a} = \kappa\s{b} = \kappa/2$. However, in this case the beam-splitter coupling is renormalized by an additional damping term i.e. $\Lambda_\mathrm{BS}\rightarrow\Lambda_\mathrm{BS}-i\kappa/4$, which prevents realizing the optimal balance of the two-mode squeezing coupling $\Lambda_\mathrm{TMS}$.

\subsubsection{Dynamical stability}
\label{A_dyn_stability}

The stability of the steady-state solution for the case of symmetrical damping rates is determined by the properties of the drift matrix in~\cref{MXP}, 
To ensure stability, all eigenvalues of the drift matrix must have negative real parts~\cite{dejesusRouthHurwitzCriterionExamination1987}, i.e., $\mathbf{Re}[\epsilon_{\pm, \pm(\mp)}] < 0$. In a symmetrical configuration, the eigenvalues can be written as
\begin{subequations}
\label{EV_MXP}
\begin{align}
\epsilon_{\pm, +} & = -\frac{\kappa}{2} \pm \sqrt{\Lambda\s{TMS}^2 - \Lambda\s{BS}^2 + 2 \left( \Lambda\s{S\s{a}}^2 + \Lambda\s{S\s{b}}^2 \right) -\frac{( \Delta\s{a}^2 + \Delta\s{b}^2 )}{2}  +  \frac{\epsilon \epsilon}{2}}, \nonumber \\
\\
\epsilon_{\pm, -} & = -\frac{\kappa}{2} \pm \sqrt{\Lambda\s{TMS}^2 - \Lambda\s{BS}^2 + 2 \left( \Lambda\s{S\s{a}}^2 + \Lambda\s{S\s{b}}^2 \right) -\frac{( \Delta\s{a}^2 + \Delta\s{b}^2 )}{2}  - \frac{\epsilon \epsilon}{2}}, \nonumber
\end{align}
\end{subequations}
where 
\begin{align}
\label{ee}
\epsilon\epsilon & \equiv \Big\lbrace \Big[ (\Delta\s{a}^2-\Delta\s{b}^2) - 4 \left(\Lambda\s{S\s{a}}^2-\Lambda\s{S\s{b}}^2 \right) \Big]^2 + 4 \Lambda\s{BS}^2 \Big[ (\Delta\s{a}+\Delta\s{b})^2 \nonumber \\
& - 4 (\Lambda\s{S\s{a}}-\Lambda\s{S\s{b}})^2 \Big] - 4 \Lambda\s{TMS}^2 \Big[ (\Delta\s{a}-\Delta\s{b})^2 - 4 (\Lambda\s{S\s{a}} + \Lambda\s{S\s{b}})^2 \Big] \nonumber \\ 
& + 32 \Lambda\s{BS} \:\Lambda\s{TMS} \left( \Lambda\s{S\s{a}} \:\Delta\s{b} + \Lambda\s{S\s{b}} \:\Delta\s{a} \right) \Big\rbrace^{1/2}. 
\end{align}
When $\Lambda\s{S\s{a}} = \Lambda\s{S\s{b}} = \Lambda\s{S}$ and $\Delta\s{a} = -\Delta\s{b} = -2\Lambda\s{S}$, and by defining the cooperativity of each coupling parameter as $\mathcal{C}_{\alpha} = 4|\Lambda_{\alpha}|^2 / \kappa^2$ with $\alpha \in \lbrace \mathrm{TMS}, \mathrm{BS}, \mathrm{S} \rbrace$, the eigenvalues can be further simplified as
\begin{align}
\label{Eigenvalues_M2}
\epsilon_{\pm, \pm} & = \frac{\kappa}{2} \left(-1 \pm \sqrt{\mathcal{C}_{\mathrm{TMS}} - \mathcal{C}_{\mathrm{BS}}} \right). 
\end{align}
As discussed in the main text, our analysis focuses on two characteristic operating points: the exceptional point (EP) and the Bogoliubov point (BP), where the system remains dynamically stable. The EP arises when when $\mathcal{C}\s{TMS} = \mathcal{C}\s{BS}$, whereas the BP is defined by $\mathcal{C}\s{TMS}$~-~$\mathcal{C}\s{BS}$~=~-1, corresponding to equal magnitudes of the real and imaginary components of the eigenvalues. As shown in~\cref{fig_phase_diagram}(a) of the main text, moving away from the instability region and approaching the BP leads to a broader BW. Further increasing $\mathcal{C}\s{BS}$ beyond the BP, results in normal-mode splitting which manifests as a splitting of the gain profiles.
\begin{figure}[!t]
\includegraphics[width = 1\columnwidth]{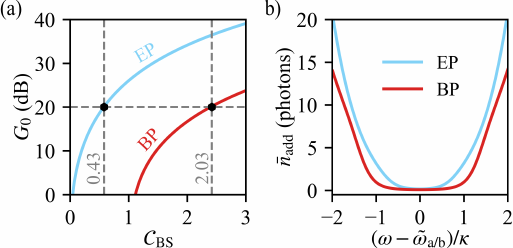}
\caption{\textbf{Comparison of predicted gain and added noise at the EP and BP.}
(a) Zero-frequency gain $\mathcal{G}\s{0}$ as a function of the beam-splitter cooperativity $\mathcal{C}\s{BS}$ for both the EP (blue) and BP (red). The black dots indicate the operating points corresponding to a gain of $\mathcal{G}\s{0} = 20~\mathrm{dB}$ at $\mathcal{C}\s{BS} = 0.58$ and $2.42$ for the EP and BP, respectively.
(b) Added noise $\bar{n}\s{add}$ as a function of normalized frequency $(\omega-\tilde{\omega}\mathrm{_{a/b}})/\kappa$ for the same operating points indicated by the black dots in panel (a). While both configurations can reach the quantum limit, at the BP we maintain near-quantum-limited performance over a broader frequency range, stemming from its enhanced bandwidth.}
\label{fig_Gain_coop_noise_theory}
\end{figure}
\subsubsection{Predicted gain and added noise}
\label{A_Gain_noise}
To extract the gain and added noise of the amplifiers, we employ the input–output formalism~\cite{gardinerQuantumNoiseHandbook2004a} in the frequency domain
\begin{equation}
\label{Inout_X}
\hat{\vec{X}}\s{out} (\omega) =  \hat{\vec{X}}\s{in} (\omega) + \mathbf{K}\s{ext} \:  \hat{\vec{X}} (\omega),
\end{equation}
where $\hat{\vec{X}} (\omega)= \mathbf{M}^{-1} (\omega) \: \mathbf{K}\s{xp} \: \hat{\Vec{X}}\s{in} (\omega)$, and $\mathbf{K}\s{ext} = \mathbf{K}\s{xp}$. Therefore, the scattering matrix, which connects the output and input quadratures through $\hat{\vec{X}}\s{out} (\omega) = \mathbf{S}(\omega) \: \hat{\vec{X}}\s{in} (\omega)$, is given by
\begin{equation}
    \label{SM}
    \mathbf{S}(\omega) = \mathbf{K}\s{xp} \: \mathbf{M}^{-1}(\omega) \: \mathbf{K\s{xp}} + \mathbf{1}\s{\mathrm{4}}.
\end{equation}
For symmetric couplings of the grAl resonators, the scattering matrix at zero frequency can be expressed in the simplified form
\begin{align}
\label{S0}
   \mathbf{S}[0] = \begin{pmatrix}
s\s{11} & 0 & 0 & s\s{14}
\\
s\s{21} & s\s{11} & s\s{23} & s\s{24} 
\\
s\s{24} & s\s{14} & s\s{11} & s\s{34} 
\\
s\s{23} & 0 & 0 & s\s{11}
\end{pmatrix},
\end{align}
with
\begin{subequations}
\begin{align}
\label{S0_elements}
s\s{11} & = \frac{-1+\mathcal{C}\s{BS}-\mathcal{C}\s{TMS}}{1 + \mathcal{C}\s{BS}-\mathcal{C}\s{TMS}}, \\
s\s{14} & = \frac{2 \left(\sqrt{\mathcal{C}\s{BS}}-\sqrt{\mathcal{C}\s{TMS}} \right)}{1 + \mathcal{C}\s{BS}-\mathcal{C}\s{TMS}}, \\
s\s{23} & = - \frac{2 \left(\sqrt{\mathcal{C}\s{BS}}+\sqrt{\mathcal{C}\s{TMS}} \right)}{1 + \mathcal{C}\s{BS}-\mathcal{C}\s{TMS}}, \\
s\s{24} & = \frac{16 \sqrt{\mathcal{C}\s{S}} \, \sqrt{\mathcal{C}\s{TMS}}}{\left(1 + \mathcal{C}\s{BS}-\mathcal{C}\s{TMS} \right)^2}, \\
s\s{21} & = \frac{8 \sqrt{\mathcal{C}\s{S}} \, \left(1 + \left(\sqrt{\mathcal{C}\s{BS}} + \sqrt{\mathcal{C}\s{TMS}} \right)^2 \right)}{\left(1 + \mathcal{C}\s{BS}-\mathcal{C}\s{TMS} \right)^2}, \\
s\s{34} & = \frac{8 \sqrt{\mathcal{C}\s{S}} \, \left(1 + \left(\sqrt{\mathcal{C}\s{BS}} - \sqrt{\mathcal{C}\s{TMS}} \right)^2 \right)}{\left(1 + \mathcal{C}\s{BS}-\mathcal{C}\s{TMS} \right)^2}.
\end{align}
\end{subequations}
The maximum gain of the amplifier at zero frequency is then given by
\begin{align}
\label{G0XP}
G\s{0} =  \Big|s\s{21} (\omega=0) \Big|^2 = \Big| \frac{8 \, \sqrt{\mathcal{C}\s{S}} \, \left( 1 + \left(\sqrt{\mathcal{C}\s{BS}} + \sqrt{\mathcal{C}\s{TMS}} \right)^2 \right)}{\left( 1 + \mathcal{C}\s{BS} - \mathcal{C}\s{TMS} \right)^2} \Big|^2.
\end{align}
In~\cref{fig_Gain_coop_noise_theory}(a), the zero-frequency gain is shown as a function of the BS coupling for both the EP and BP, assuming $\mathcal{C}\s{S} = \mathcal{C}\s{TMS}/2 $. The BP configuration exhibits a broader bandwidth at 20~dB gain, which is obtained at the expense of increased BS cooperativity.

We calculate the added noise at both EP and BP using the symmetrized noise spectral density.
\begin{equation}
    \label{noise_power}
    \bar{S}\s{F_\textit{i},F_\textit{i}}^{\mathrm{(out)}} (\omega) = \frac{1}{2} \int \frac{d \Omega}{2 \pi} \: \left\langle \left\lbrace \hat F\s{\textit{i},out}(\omega), \hat F\s{\textit{i},out}(\Omega) \right\rbrace \right\rangle, 
\end{equation}
where $\hat F\s{\textit{i},out} = \lbrace \hat X\s{\textit{i},out} , \hat P\s{\textit{i},out} \rbrace$ with $ i = \mathrm{a},\mathrm{b}$. The amplifier added noise photons are then calculated using
\begin{equation}
    \label{naddedw}
    \bar{n}\s{add} (\omega) = \frac{\bar{S}^{\mathrm{(out)}}\s{\hat{P}\s{a},\hat{X}\s{a}} (\omega)}{G(\omega)} - \frac{1}{2},
\end{equation}
where $G(\omega)=|s_\mathrm{21}(\omega)|^2$. In~\cref{fig_Gain_coop_noise_theory}(b), we show the calculated added noise photons associated with the quadrature $\hat{P}\s{a,out}$ for the amplifier operating at both the EP and BP. The added noise for other quadratures is expected to exhibit a similar trend. For both the EP and BP, our calculations predict the realization of a broadband quantum-limited amplifier. Notably, the BP amplifier sustains a lower added-noise level over a broader frequency range.

\subsubsection{Dynamics in presence of a single pump}
\label{A_Theory_Single_Pump_hybrid}
Under single pump operation, the system Hamiltonian in the hybridized basis becomes
\begin{align}
   \label{H_lin_H_single}
    \hat H\s{H} / \hbar & = \sum_{i = \mathrm{a}, \mathrm{b}} \Big \lbrace \tilde{\omega}_i \: \hat c_i^{\dagger} \hat c_i  + \Big( \Lambda\s{S_\textit{i}} \: \mathrm{e}^{-2 i \omega\s{g} t}  \: \hat c_i^{\dagger} \hat c_i^{\dagger} + \mathrm{h.c} \Big) \Big \rbrace  \\
    & + \Big( \Lambda\s{TMS} \: \mathrm{e}^{-2 i \omega\s{g} t} \: \hat c_a^{\dagger} \hat c_b^{\dagger} + \mathrm{h.c} \Big) + \Lambda\s{BS} \left(\hat c_a^{\dagger} \hat c_b + \hat c_b^{\dagger} \hat c_a \right), \nonumber
\end{align}
where
\begin{align}
\label{WAB_HL_single}
\tilde{\omega}\s{a}  & = \omega\s{a}  + 2 \mathcal{K}\s{L} \, |\alpha\s{L}|^2 \sin^2 \theta + 2 \mathcal{K}\s{R} \, |\alpha\s{R}|^2  \cos^2\theta , \nonumber \\
\\
\tilde{\omega}\s{b}   & = \omega\s{b}  +  2 \mathcal{K}\s{L} \, |\alpha\s{L}|^2 \cos^2 \theta + 2 \mathcal{K}\s{R} \, |\alpha\s{R}|^2 \sin^2\theta  , \nonumber 
\end{align}
and
\begin{align}
\label{WAB_HL_single}
\Lambda\s{S\s{a}} & =\mathcal{K}\s{L} \alpha\s{L}^2 \sin^2 \theta + \mathcal{K}\s{R} \alpha\s{R}^2 \cos^2 \theta,  \nonumber \\
\\
\Lambda\s{S\s{b}} & =\mathcal{K}\s{L} \alpha\s{L}^2 \cos^2 \theta + \mathcal{K}\s{R} \alpha\s{R}^2 \sin^2 \theta,  \nonumber
\end{align}
\begin{align}
\label{WAB_HL_single}
\Lambda\s{TMS} & = -\frac{1}{2} \left( \mathcal{K}\s{L} \alpha\s{L}^2 - \mathcal{K}\s{R} \alpha\s{R}^2 \right) \sin 2\theta,  \\
\nonumber \\
\Lambda\s{BS} & = - \left( \mathcal{K}\s{L} |\alpha\s{L}|^2 - \mathcal{K}\s{R} |\alpha\s{R}|^2 \right) \sin 2\theta.
\end{align}
In the frame rotating at $\omega\s{g}$, the Hamiltonian in the hybridized basis takes the form
\begin{align}
   \label{H_lin_H_single_R}
    \hat H\s{H} / \hbar & = - \sum_{i = \mathrm{a}, \mathrm{b}} \Big \lbrace \Delta_i \: \hat c_i^{\dagger} \hat c_i  - \Big( \Lambda_{S_i} \: \hat c_i^{\dagger} \hat c_i^{\dagger} + \mathrm{h.c} \Big) \Big \rbrace  \\
    & + \Big( \Lambda\s{TMS} \: \hat c\s{a}^{\dagger} \hat c\s{b}^{\dagger} + \mathrm{h.c} \Big) + \Lambda\s{BS} \left(\hat c\s{a}^{\dagger} \hat c\s{b} + \hat c\s{b}^{\dagger} \hat c\s{a} \right), \nonumber
\end{align}
where $\Delta_i = \omega\s{g} - \tilde{\omega}_{i}$ and $i = \mathrm{a},\mathrm{b}$.
For a nondegenerate amplifier, the pump frequency satisfies $2 \omega\s{g} = \tilde{\omega}\s{a} + \tilde{\omega}\s{a}$. In this case, the dominant interaction terms correspond to the TMS coupling, allowing the system dynamics to be reduced to
\begin{align}
\label{H_Amp_nondeg_single}
\hat H\s{H} / \hbar \approx  -\Delta\s{a}  \: \hat c\s{a} ^{\dagger} \hat c\s{a}  - \Delta\s{b}   \: \hat c\s{b} ^{\dagger} \hat c\s{b}  +\left(\Lambda\s{TMS} \: \hat c\s{a}^{\dagger} \hat c\s{b}^{\dagger} + \mathrm{h.c.} \right),
\end{align}
which, in the quadrature representation, takes the form
\begin{align}
\label{H_Amp_nondeg_single_xp}
\hat H\s{H} / \hbar \approx  - \sum_{i = \mathrm{a}, \mathrm{b}}  \frac{\Delta_i}{2} \left( \hat X_{i}^2 + \hat P_{i}^2 \right) + \Lambda\s{TMS} \left( \hat X\s{a} \hat X\s{b} - \hat P\s{a} \hat P\s{b} \right).
\end{align}
Therefore, for the single-pump configuration, the system operates as a phase-preserving amplifier with the maximum gain at zero frequency, which is given by
\begin{align}
\label{GainS_hyb}
G_0 = \left| \frac{\mathcal{C}\s{TMS} +1 }{\mathcal{C}\s{TMS} - 1} \right|^2.
\end{align}
According to this relation, the highest amplification is achieved as $\mathcal{C}\s{TMS} \rightarrow 1$. However, this condition is limited by the dynamical stability requirement $\mathcal{C}\s{TMS} < 1$ (see the denominator of~\cref{GainS_hyb}), which imposes the GBW scaling in the single-pumped grAlPA.

\begin{figure*}[t!]
\includegraphics[width=6.67in]{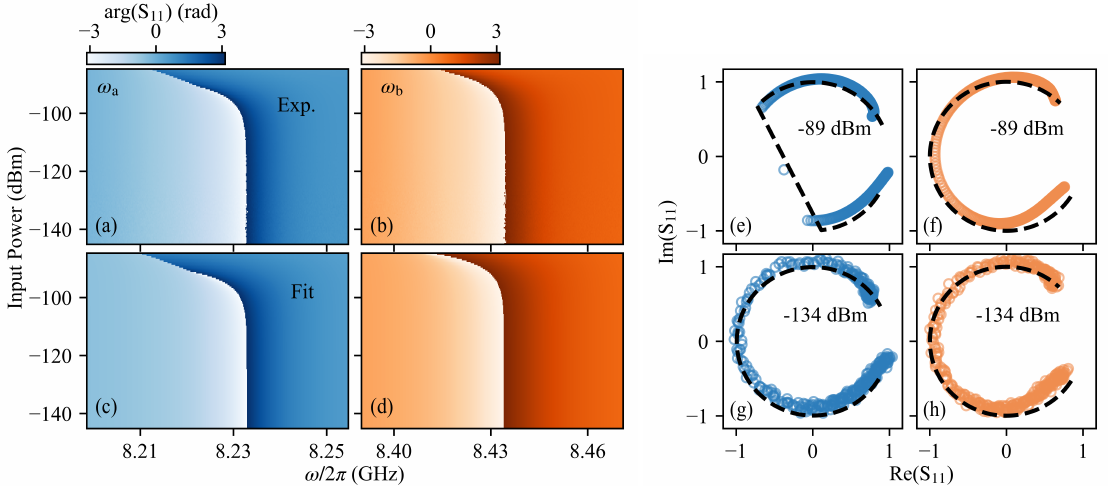}
\caption{\textbf{Kerr-shift of dimer modes and extraction of circuit parameters.} Measured~\textbf{(a)}-\textbf{(b)}~and~fitted~\textbf{(c)}-\textbf{(d)}~single-tone spectroscopy of the grAlPA dimer modes $\omega\mathrm{_{a/b}}$ as a function of input power. Fits are obtained using Eqs.(\ref{eq_Gamma_dimer})-(\ref{eq_photons_dimer}) with the circuit parameters from \cref{fig_design}(a) as variables. The probe power is calibrated from the measurement-induced dephasing of a superconducting qubit placed before the amplifier (see \cref{A_PowerCal}). Right-hand panels show linecuts at probe powers of –89~dBm (\textbf{e}-\textbf{f}) and –134~dBm (\textbf{g}-\textbf{h}). The black dashed lines depict the fit curves.}
\label{fig_Kerr_fit}
\end{figure*}
\section{Calculation of circuit parameters}
\label{A_circuitParameters}
We extract the grAlPA circuit parameters from the power dependent single-tone spectroscopy measurements shown in \cref{fig_Kerr_fit}(a)-(b). First, we obtain the dimer modes frequencies $\omega\mathrm{_{a/b}}$ and linewidths $\kappa\mathrm{_{a/b}}$ by performing circle fits of both resonances close to the single-photon regime ($\sim$ -140~dBm input power), where we can neglect the effect of the self-Kerr coefficients $\mathcal{K}\mathrm{_{L/R}}$.  Afterwards, we calculate $\omega\mathrm{_{L/R}}$, $J$ and $\kappa$ from \cref{eq_hybridized_frquencies} and the following expression \cite{Eichler2014Dimer}

\begin{equation}
\kappa\mathrm{_{a/b}} =\frac{\kappa}{2}\left( 1 \pm \frac{\omega\mathrm{_L}-\omega\mathrm{_R}}{\sqrt{4J^2 + \left(\omega\mathrm{_L}-\omega\mathrm{_R}\right)^2}} \right).
\label{eq_kappas_omegas_hybrid}
\end{equation}

At higher input powers, the standard circle-fit procedures fails to reproduce the correct resonance lineshapes, necessitating instead the generalized formula of the dimer reflection coefficient to calculate $\mathcal{K}\mathrm{_{L/R}}$. Following the approach of Refs. \cite{Eichler2014Sat,Eichler2014Dimer}, we derive the reflection coefficient $\Gamma$ as function of frequency $\omega$

\begin{table}[t]
\caption{Circuit parameters of grAlPA extracted from the fits presented in \cref{fig_Kerr_fit}. 
We use the estimate $\gamma/2\pi \leq$~0.1~MHz \cite{Zapata2024Dec}. 
Error intervals arise from the Fano uncertainty in the measurement setup \cite{Rieger2023fano}.}
\label{tab_Circuit_parameters}
\begin{center}
\begin{tabular}{ l | r }
\hline \\[-0.25cm]
 $\omega_a/2\pi$ (GHz) &   8.233 \\[0.12cm]
 $\omega_b/2\pi$ (GHz) &   8.434 \\[0.12cm]
 $\kappa_a/2\pi$ (MHz) &  $14 \pm 1$ \\[0.12cm]
 $\kappa_b/2\pi$ (MHz) &  $29 \pm 3$ \\[0.12cm]
 $\omega_L/2\pi$ (GHz) &  $8.299 \pm 0.007$ \\[0.12cm]
 $\omega_R/2\pi$ (GHz) &  $8.368 \pm 0.007$ \\[0.12cm]
 $J/2\pi$ (MHz)        &  $95 \pm 5$ \\[0.08cm]
 $\kappa/2\pi$ (MHz)   &  $44 \pm 4$ \\[0.08cm]
 $\gamma/2\pi$ (MHz)   &  $\leq 0.1$ \\[0.12cm]
 $\mathcal{K}\mathrm{_L}/2\pi$ (kHz)      &  $2.9 \pm 1$ \\[0.08cm]
 $\mathcal{K}\mathrm{_R}/2\pi$ (kHz)      &  $3.2 \pm 1$ \\[0.08cm]
\hline
\end{tabular}
\end{center}
\end{table}
\begin{equation}
\Gamma\left(\omega\right)=1+\frac{i\kappa\left(\delta\mathrm{_R}-i\gamma/2\right)}{J^2 - \left(\delta\mathrm{_R}-i\gamma/2\right)\left(\delta\mathrm{_L}-i\kappa/2\right)},
\label{eq_Gamma_dimer}
\end{equation}
where $\delta_j=\omega-\omega_j-\mathcal{K}_j\,n\mathrm{_{j}}$ (with $j$~=~$\mathrm{L,R}$) is the Kerr-shifted detuning from the bare grAl frequencies $\omega\mathrm{_{L/R}}$, $n\mathrm{_{L/R}}=|\alpha\mathrm{_{L,R}}|^2$ their respective photon number populations and $\gamma$ the effective damping rate due to internal losses, previously reported to lie in the sub-MHz range \cite{Zapata2024Dec}. Using the steady-state Langevin equations of the bare grAl modes (cf.~\cref{eq_langevin_bare_basis}), we obtain expressions for $n\mathrm{_{L/R}}$

\begin{subequations}
\begin{align}
n\mathrm{_{R}} & = n\mathrm{_{L}}\,\frac{J^2}{\delta\mathrm{_R}^2+\gamma^2/4}
\\
n\mathrm{_{L}}  & = n\mathrm{_{in}}\,\frac{\delta\mathrm{_R}^2+\gamma^2/4}{\left(J^2 - \delta\mathrm{_R}\delta\mathrm{_L}-\gamma\kappa/4\right)^2+\left(\gamma\,\delta\mathrm{_L}/4+\kappa\,\delta\mathrm{_R}/4\right)^2},
\end{align}
\label{eq_photons_dimer}
\end{subequations}
where $n\mathrm{_{in}}$ can be calculated from the power at the amplifier input $P\mathrm{_{in}}$ (see \cref{A_PowerCal} for details about the power calibration) using $n\mathrm{_{in}}=P\mathrm{_{in}}/\hbar\omega\kappa$. Finally, we fit the spectroscopy data beyond the single-photon regime by solving equations (\ref{eq_Gamma_dimer})-(\ref{eq_photons_dimer}) self-consistently and taking only $\mathcal{K}\mathrm{_{L/R}}$ as fitting parameters. The results are illustrated in \cref{fig_Kerr_fit}(c)-(h) and the final circuit variables are summarized in \cref{tab_Circuit_parameters}. Notably, with our procedure we can reproduce the lineshape of mode $\omega\mathrm{_a}$ even beyond its bifurcation point (see \cref{fig_Kerr_fit}(e)).

\section{Measurement setup}
\label{A_Setup}

\begin{figure}[t]
\begin{center}
\includegraphics[width = 1\columnwidth]{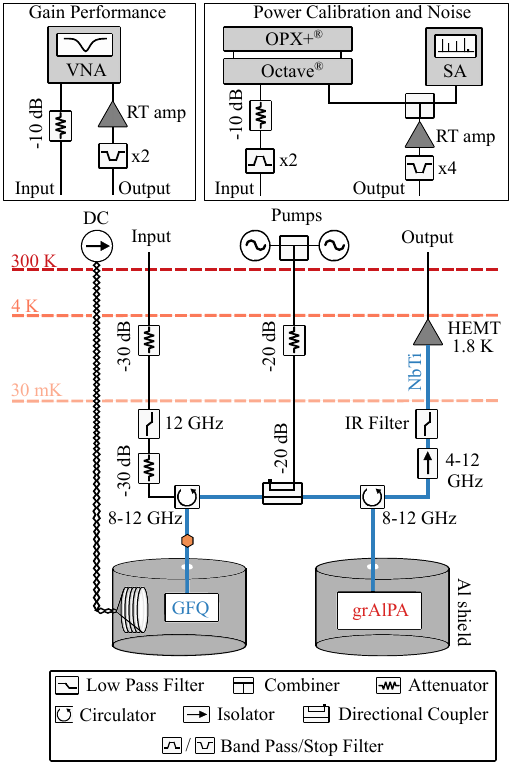}
\caption{\textbf{Experimental setups used for gain measurements, and for power calibration and noise performance.} Each component is thermalized at the temperature stage indicated by the nearest dashed line. Gain performance is characterized using a Vector Network Analyzer (VNA). For the power calibration described in Appendix~\ref{A_PowerCal} and for grAlPA noise measurements, we employ a combination of Quantum-Machines OPX-Octave$^{\text{\textregistered}}$ electronics and a Spectrum Analyzer (SA). Pump tones are generated by independent microwave sources, except for the phase-sensitive gain measurements of \cref{fig_phase_dependent_gain}, where they are synthesized directly by the OPX-Octave$^{\text{\textregistered}}$ system.}
\label{fig_Wiring}
\end{center}
\end{figure}

In our experiments, we use the two setups depicted in \cref{fig_Wiring}. For the grAlPA low power characterization (see \cref{A_circuitParameters}) and gain performance measurements (see \cref{fig_gain_protocol} and \cref{A_Gain Fits}) we use a Vector Network Analyzer (VNA) to measure the transmission between the input and output lines of the cryostat. A probe tone generated by the VNA is attenuated by a chain consisting of a -10~dB attenuator at 300~K and two -30~dB attenuators at 4~K and 30~mK, respectively, alongside a 12~GHz cut-off low-pass filter located at the mixing chamber of the refrigerator. The signal is then routed through a Generalized Flux qubit (GFQ) and the grAlPA via two cryogenic circulators, before passing through a two-stage isolator and an infrared (IR) filter \cite{HERD}. The output is subsequently amplified by a HEMT amplifier thermalized at 4~K \cite{HEMT_LNF} and a commercial room-temperature amplifier. For all connections between the GFQ, the grAlPA and the HEMT we employ NbTi superconducting cables, which minimize insertion loss and reduce uncertainty in the power calibration. Pump tones generated by independent microwave sources are combined at room temperature using a Wilkinson power combiner, and subsequently merged with the probe tone at millikelvin temperatures via a directional coupler. To suppress pump-tone leakage to the room temperature electronics, we employ a series of homemade notch filters with sub-1~dB insertion loss in the pass-band. 

For the power calibration using the dephasing of the GFQ (see \cref{A_PowerCal}) and the noise performance of the grAlPA (see \cref{fig_noise_performance}) we employ a combination of time-domain and frequency-domain measurements with an OPX-Octave$^{\text{\textregistered}}$ system \cite{Octave} and a Spectrum Analyzer (SA). Two additional notch filters are included to further protect the OPX-Octave$^{\text{\textregistered}}$ from the pump drives. For the phase-dependent gain measurements shown in \cref{fig_phase_dependent_gain}, the probe and the two pump tones are synthesized from the same OPX-Octave$^{\text{\textregistered}}$ output line. To suppress unwanted spurious tones in the generated signals, we use two homemade band-pass filters centered at both pump frequencies.

\section{Gain profiles close to each dimer mode~$\omega\mathrm{_{a,b}}$}
\label{A_Full gain profiles}

\begin{figure*}[t!]
\includegraphics[width=6.67in]{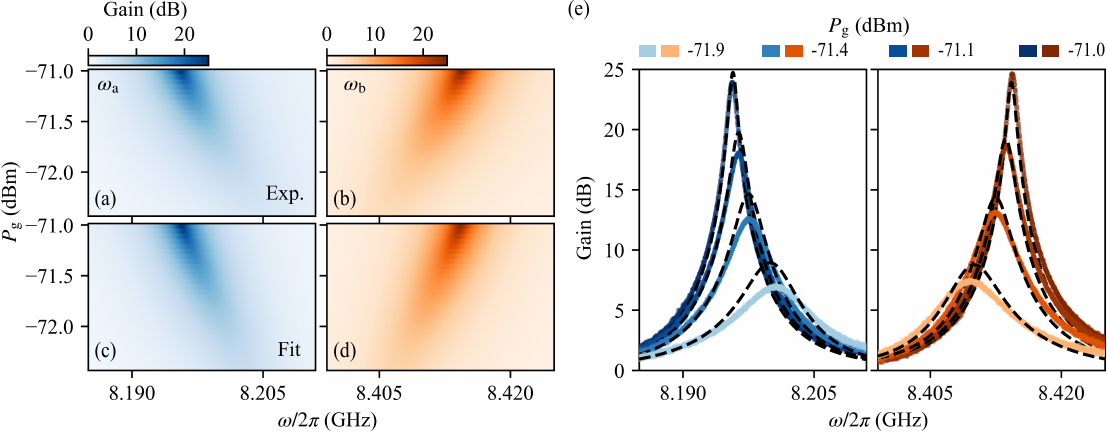}
\caption{\textbf{Power dependent gain profiles for the single-pump grAlPA.} Measured~\textbf{(a)}-\textbf{(b)}~and~fitted~\textbf{(c)}-\textbf{(d)}~gain profiles of the grAlPA dimer modes $\omega\mathrm{_{a/b}}$ as a function of $P\mathrm{_g}$. We keep $\omega\mathrm{_g}$ fixed. Fits are obtained using the procedure explained in \cref{A_Gain Fits} with the pump line attenuation as the only fitting parameter. \textbf{(e)} Linecuts at different pump powers. The black dashed lines are the fitted curves.}
\label{fig_Single_Pump_Fit}
\end{figure*}

\begin{figure*}[t!]
\includegraphics[width=6.67in]{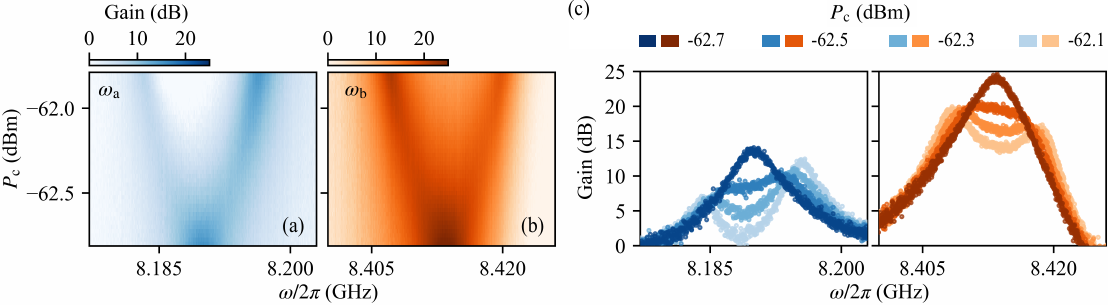}
\caption{\textbf{Power dependent gain profiles for the double-pump grAlPA.} Measured~\textbf{(a)}-\textbf{(b)}~gain profiles of the grAlPA dimer modes $\omega\mathrm{_{a/b}}$ as a function of $P\mathrm{_c}$. We keep $\omega\mathrm{_g}$, $\omega\mathrm{_c}$ and $P\mathrm{_g}$ fixed. \textbf{(e)}~Linecuts at different pump powers.}
\label{fig_Double_Pump_Fit}
\end{figure*}

In \cref{fig_Single_Pump_Fit} and \cref{fig_Double_Pump_Fit}, we present examples of measured gain profiles near $\omega\mathrm{_{a/b}}$ for a single-pumped and double-pumped grAlPA, respectively. Color maps in \cref{fig_Single_Pump_Fit}(c)-(d), and the black dashed lines in \cref{fig_Single_Pump_Fit}(d) are obtained from fits using the Bose-Hubbard dimer model (cf. \cref{A_Gain Fits}). The BW scaling follows the relation GBW$= \kappa\mathrm{_{eq}}/G\mathrm{_0}$, where $G\mathrm{_0}$ is the maximum gain and $\kappa\mathrm{_{eq}}=2\kappa\mathrm{_a}\kappa\mathrm{_b}/(\kappa\mathrm{_a}+\kappa\mathrm{_b})$ is the equivalent damping rate~\cite{Eichler2014Dimer}. The observed asymmetry of the gain profiles close to $\omega\mathrm{_{a}}$ and $\omega\mathrm{_{b}}$ originates from the asymmetry in the coupling of the bare grAl resonators.  

\section{Fits of Gain Profiles based on Bose-Hubbard dimer model}
\label{A_Gain Fits}
We fit the experimental gain profiles using the results of the multi-pumped Bose-Hubbard dimer model shown in \cref{A_Theory} and the parameters extracted in \cref{A_circuitParameters}. For the single-pump grAlPA, we calculate first the average pump fields $\alpha\mathrm{_{L/R}}$ from the mean-field Langevin equations \cref{eq_langevin_bare_basis}. Since the amplifier operates in a multistable regime, multiple solutions of $\alpha\mathrm{_{L/R}}$ exist for the same input field $\alpha\mathrm{^g_{in}}$. We select the solution corresponding to the lowest photon population in the grAl resonators. The frequency-dependent gain is then obtained using the formula $G=|\mathrm{S}\mathrm{_{00}}(\omega)|^2$, where $\mathbf{S}(\omega)$ is defined in \cref{SM_B}. This procedure is repeated iteratively, using only the pump-line attenuation as a fitting parameter. The results, shown in \cref{fig_Single_Pump_Fit}, are consistent with a pump-line attenuation of -66.4~$\pm$~5~dB.  The observed asymmetry of the gain profiles and the discrepancy at low pump powers can be attributed to unaccounted frequency-dependent losses in the measurement setup.

\begin{table}[b]
\caption{Fitted pump powers for the grAlPA under optimal double-pump. We use the gain profiles in \cref{fig_gain_protocol}(c) to determine $G\mathrm{_0}$.}
\label{tab_powers_double_pump}
\begin{center}
\begin{tabular}{  c | c c c c } 
\hline 
 & \multicolumn{4}{| c}{Maximum gain $G\mathrm{_0}$} \Tstruttop\Bstruttop \\[0.11cm]
\hline
\hfil Quantity & \hfil 10 dB & \hfil 15 dB & \hfil 20 dB & \hfil 25 dB 
\\ 
\hline
\hfil $P\mathrm{_g}$ (dBm) & \hfil -71.50 & \hfil -71.54 & \hfil -71.76 & \hfil -71.93 \Tstrut\Bstrut \\ 
\hfil $P\mathrm{_c}$ (dBm) & \hfil -62.61 & \hfil -62.64 & \hfil -62.50 & \hfil -62.40 \Tstrut\Bstrut \\ 
\hfil $P\mathrm{_c}/P\mathrm{_g}$ & \hfil 0.13 & \hfil 0.13 & \hfil 0.12 & \hfil 0.11 \Tstrut\Bstrut \\ 
\hline
\end{tabular}
\end{center}
\end{table}

Under two-pump operation, each gain profile shown in~\cref{fig_gain_protocol}(c) is fitted independently by solving \cref{eq_steady_state_Langevin_alphas} and \cref{SM_B} self-consistently and using the pump tones powers as free parameters. We calculate the frequency-dependent gain similar to the single-pumped grAlPA with \cref{SM_B}. We iteratively adjust the resonators mean-field amplitudes $\alpha_\mathrm{L,R}$ until the simulated profiles match the experimental data. Finally, we extract the equivalent pump powers using \cref{eq_steady_state_Langevin_alphas}. In~\cref{tab_powers_double_pump} we summarize the calculated pump powers corresponding to each gain profile in~\cref{fig_gain_protocol}(c). All values are within the uncertainty range of the pump-line attenuation calibration obtained with a single pump. Moreover, the ratio between the conversion and gain pump powers remains approximately constant for different gain levels.

\section{Power calibration}
\label{A_PowerCal}

\begin{figure}[t]
\begin{center}
\includegraphics[width = 1\columnwidth]{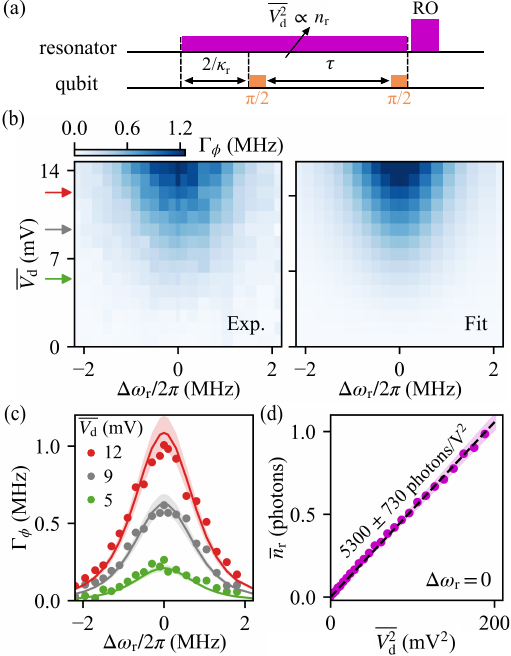}
\caption{\textbf{Measurement-induced dephasing and power calibration using a Generalized Flux Qubit.} \textbf{(a)}~Pulse sequence used to obtain the measurement-induced dephasing $\Gamma\mathrm{_{\phi}}$ of the GFQ using Ramsey interferometry. To reach steady-state, the resonator is populated $2/\kappa\mathrm{_r}$~=~125~ns before the start of the qubit manipulation. We use a readout (RO) pulse with a power level equivalent to 20 measurement photons. \textbf{(b)}~Measured (left) and fitted (r) qubit dephasing as a function of readout voltage $\overline{V\mathrm{_{d}}}$ and readout detuning $\Delta\mathrm{_r}$.~\textbf{(c)}~Linecuts of $\Gamma\mathrm{_{\phi}}$ for three different readout voltages. The solid lines correspond to the resulting fits using \cref{eq_Gamma_detun}. \textbf{(d)}~Resonator occupation number $\overline{n}_\mathrm{_r}$ as a function of $\overline{V\mathrm{^2_{d}}}$ at $\Delta\omega\mathrm{_r}$~=~0. The black dashed line corresponds to the resulting fit using \cref{eq_photons_Gamma}.}
\label{fig_Power_cal}
\end{center}
\end{figure}

We calibrate the input-line attenuation using the measurement-induced dephasing of a Generalized Flux Qubit (GFQ) in a circuit QED setup \cite{Geisert2024Aug}. We start by using single-tone spectroscopy to extract the frequency $\omega\mathrm{_r}/2\pi$~=~8.060~GHz and linewidth $\kappa\mathrm{_r}/2\pi$~=~2.54~$\pm$~0.26~MHz of a readout resonator coupled to the GFQ. Next, we perform Ramsey interferometry experiments while driving the resonator with a room temperature voltage amplitude $V\mathrm{_d}$, as illustrated in \cref{fig_Power_cal}(a). The drive tone induces an average photon population $\bar{n}\mathrm{_r}$ and a qubit dephasing rate $\Gamma_{\phi}$ given by \cite{Gambetta2006Oct} 

\begin{equation}
\begin{aligned}
\bar{n}\mathrm{_r} & = c\overline{V\mathrm{^2_{d}}}\left(\frac{\kappa^2+\chi^2}{\kappa^2+ \left(2\Delta\omega\mathrm{_r}
 +\chi\right)^2}+\frac{\kappa^2+\chi^2}{\kappa^2+\left(2\Delta\omega\mathrm{_r}-\chi\right)^2}\right),
\end{aligned}
\label{eq_photons_Gamma}
\end{equation}

and

\begin{equation}
\Gamma_{\phi}  = \frac{\bar{n}\mathrm{_r} \kappa\chi^2}{\kappa^2+\chi^2+\left(2\Delta\omega\mathrm{_r}\right)^2},
\label{eq_Gamma_detun}
\end{equation}
where $\Delta\omega\mathrm{_r}=\omega\mathrm{_d}-\omega\mathrm{_r}$ is the frequency detuning of the driving tone, $\chi$ is the dispersive shift of the GFQ and $c$ a proportionality constant relating $\overline{V\mathrm{^2_{d}}}$ and the power at the input of the readout resonator. We calculate $\Gamma_{\phi}$ as a function of $\Delta\omega\mathrm{_r}$ and $V\mathrm{_d}$ from the measured Ramsey fringes and we fit the data using Eqs.~(\ref{eq_photons_Gamma})-(\ref{eq_Gamma_detun}) with $\chi$ and $c$ as fitting parameters. Figures \ref{fig_Power_cal}(b)-(d) show the fitting results, from which we obtain $\chi/2\pi$~=~0.51~$\pm$~0.05~MHz and $c$~=~5300~$\pm$~730 photons/V$\mathrm{^2}$ corresponding to an input-line attenuation of -112.5~$\pm$~0.5~dB. 

Notice that the power calibration is valid at the input of the GFQ readout resonator. To refer the power to the grAlPA input, we account for the total insertion loss between the GFQ and the amplifier, for which we estimate an upper bound of 0.6~dB. In addition, since the readout frequency $\omega\mathrm{_r}$ is detuned by approximately 200~MHz from the dimer modes, we expect an additional uncertainty arising from ripples in the microwave connections. Following the results of Ref. \cite{Zapata2024Dec}, obtained in a similar setup, we assign a 1~dB uncertainty to the final power calibration. The error bars in \cref{fig_noise_performance} are obtained from the propagation of this uncertainty. For the setup used in gain measurements (see \cref{A_Setup}), we calculate the change in attenuation by accounting for the insertion loss of the components connected outside the cryostat.

\begin{figure}[b]
\begin{center}
\includegraphics[width = 1\columnwidth]{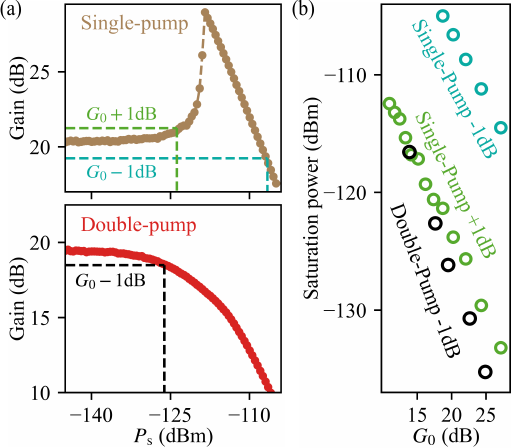}
\caption{\textbf{grAlPA saturation power.} \textbf{(a)}~Saturation power measurements when the grAlPA is driven with one (top) or two (bottom)  pumps. When operated with a single pump, the grAlPA exhibits a ‘shark-fin’-like saturation curve, marking the onset of multistable dynamics \cite{Sivak2019}. \textbf{(b)}~Saturation power scaling as a function of low-power gain $G\mathrm{_0}$. Remarkably, the double-pump –1~dB compression point closely matches the +1~dB point in the single-pump case, suggesting that the grAlPA saturation power is limited by device multistability.}
\label{fig_sat_power}
\end{center}
\end{figure}

\section{Dynamic Range}
\label{A_DynamicRange}

In Figure \ref{fig_sat_power}(a) we show saturation power measurements of the grAlPA for low-power gain $G\mathrm{_0} \approx$~20~dB. When driven by a single pump, the amplifier exhibits a nonmonotonic saturation curve for increasing probe powers $P\mathrm{_s}$. This effect, colloquially referred as shark-fin \cite{Sivak2019}, originates from operating the grAlPA in a multistable regime. In this situation, we define two characteristic powers: the conventional $G\mathrm{_0}$–1~dB compression point, and the $G\mathrm{_0}$+1~dB compression point, which quantifies the power threshold above which transitions between metastable states occur \cite{Sivak2019}. We measure -1~dB compression points close to -106~dBm for the single-pumped case (similar to previously reported values \cite{Zapata2024Dec}) and -126~dBm for the double-pumped case, respectively. Strikingly, we find that the single-pumped +1~dB point and the double-pumped –1~dB point lie within the same range. As shown in \cref{fig_sat_power}(b), we observe a similar behavior across different gain levels, indicating that the dynamic range under two-pump operation is currently limited by the multistability of the amplifier, and could be improved in future designs. 


\bibliography{GrAlPA}

\end{document}